\documentclass[pra,twocolumn,aps,amssymb]{revtex4}
\usepackage{amssymb}

\usepackage{amssymb}

\usepackage{graphicx}
\usepackage{amsmath}
\usepackage{times}

\begin{document}

\title{Ultracold atoms confined in an optical lattice plus parabolic potential: a closed-form approach}
\author{ Ana Maria Rey\footnote{Electronic address: ana.rey@nist.gov}, Guido Pupillo\footnote{Electronic address: guido.pupillo@nist.gov}, Charles W. Clark, and Carl J. Williams}
\affiliation{National Institute of Standards and Technology, Gaithersburg, MD 20899}
\date{\today}

\begin{abstract}
We discuss interacting and non-interacting one dimensional atomic
systems trapped in an optical lattice plus a parabolic potential.
We show that, in the tight-binding approximation, the
non-interacting problem is exactly solvable in terms of Mathieu
functions.  We use the analytic solutions to study the collective
oscillations of ideal bosonic and fermionic ensembles induced by
small displacements of the parabolic potential. We treat the
interacting boson problem by numerical diagonalization of the
Bose-Hubbard Hamiltonian.  From analysis of the dependence upon
lattice depth of the low-energy excitation spectrum of the
interacting system, we consider the problems of "fermionization"
of a Bose gas, and the superfluid-Mott insulator transition.  The
spectrum of the noninteracting system turns out to provide a
useful guide to understanding the collective oscillations of the
interacting system, throughout a large and experimentally relevant
parameter regime.
\end{abstract}

\maketitle
\section{Introduction}
In recent experiments \cite{Tolra,Moritz,Soferle,Paredes,Weiss},
quasi-one dimensional systems have been realized by tight
confinement of gases in two dimensions. Due to the enhanced
importance of quantum correlations as dimensionality is reduced,
such systems exhibit physical phenomena not present in higher
dimensions, such as "fermionization" of a Bose-Einstein gas.

The degree of correlation is measured by the ratio of the
interaction energy to the kinetic energy, $\gamma$. For $\gamma \ll
1$  the system is weakly interacting, and at zero temperature most
atoms are Bose-condensed. In this limit quantum correlations are
negligible and the dynamics is governed by the mean-field
Gross-Pitaevski equation. For $\gamma \gg 1$, the system is highly
correlated and becomes fermionized, in that repulsive
interactions mimic the effects of the Pauli exclusion principle.
In this "Tonks-Girardeau" regime, \cite{Girardeau,Lieb}, the low
energy excitation spectrum of a bosonic gas resembles that of a
gas of non-interacting fermions.

To date, one dimensional systems have been obtained by loading a
Bose-Einstein condensate into a two-dimensional optical lattice
which is deep enough to restrict the dynamics to one dimension.
This procedure creates an array of independent 1D tubes. In most
experiments, a weak quadratic potential is superimposed upon the
lattice, in order to confine the atoms during the loading
proccess. The combined presence of the periodic and quadratic
potentials substantially modifies the dynamics of the trapped
atoms compared to the cases when only one of the two potentials is
present, as shown both experimentally and theoretically
\cite{Pezze,Ott,Ruuska,polkovnikov0,Hooley,Rigol}.

In this paper we study both ideal and interacting bosonic systems
in such potentials. In contrast to previous studies of ideal
systems, which used numerical\cite{Ruuska} or approximate
solutions \cite{Pezze,polkovnikov0,Hooley,Rigol}, here we show
that the single-particle problem is exactly solvable in terms of
Mathieu functions\cite{AS64,Meixner,math}. We use analytic
solutions to fully characterize the energy spectrum and
eigenfunctions in the various regimes of the trapping potentials
and to provide analytic expressions for the oscillations of the
center of mass of both ideal bosons and fermions that are induced
by small displacements of the parabolic trap. These expressions
for the dipolar motion may be tested in experiments.

We further analyze the low-energy spectrum of interacting bosons
by means of exact diagonalizations of the Bose-Hubbard
Hamiltonian, and identify the conditions required for
fermionization to occur. Moreover, we show that specific changes
in the spectrum of the fermionized system can be used to describe
the characteristics of a Mott insulator state with unit filling at
the trap center.

Center-of-mass oscillations are also studied in the weakly
interacting and fermionized regimes by comparing exact numerical
solutions for the interacting system to solutions for ideal bosons
and fermions, respectively. Because fermionization occurs over a
large range of trap parameters, knowledge  of the properties of
the single-particle solutions turns out to provide useful insights
in the understanding of the complex many-body dynamics. In
addition, we numerically analyze the distribution of frequencies
pertaining the modes excited during the collective dynamics, for
different values of $\gamma$. This helps us gain qualitative
insight in the dynamics even in the intermediate regime where
interaction and kinetic energies are comparable and no mapping to
ideal gases is possible.\\

The presentation of the results is organized as follows: In Sec.II
we discuss  the analytic solutions that describe the
single-particle physics. These show the existence of two types of
behavior, according to whether the energy is dominated by site
hopping or parabolic contributions. Different asymptotic
expansions of the eigenfunctions and eigenvalues apply to these
two regimes, and can be effectively combined to describe the full
spectrum.

In Sec. III, we then apply the analytic solutions and asymptotic
expansions to the description of the collective dynamics of
non-interacting ensembles of bosons and fermions subject to a
sudden displacement of the parabolic trap. In contrast to the
well-known case of the displaced harmonic oscillator, a
non-trivial modulation of the center of mass motion occurs due to
the presence of the lattice potential. We derive explicit
expressions for the initial decay of the amplitude of the
oscillation, to which we refer as effective damping.

In Sec.IV A, the low-energy spectrum of interacting bosons is
studied as a function of the lattice depth.  In particular, we
follow the evolution of the spectrum from ideal bosonic to ideal
fermionic as the lattice deepens, by means of numerical
diagonalization of the Hamiltonian for a moderate number of atoms
and wells. We specify the necessary conditions for the formation
of a Mott state at the center of the trap, and use the Fermi-Bose
mapping to link its appearance and the reduction of number
fluctuations to the population of high-energy localized states at
the Fermi level.

Section IV B is dedicated to the study of the
collective dipole dynamics of an interacting bosonic gas  by
numerical calculations of the exact quantal dynamics. The dynamics
is studied for two different scenarios that are experimentally
realizable. The first corresponds to experiments in which the
trapping potentials are kept fixed and the interatomic scattering
length is varied (e.g. by use of a Feshbach resonance), Sec.IV
B1. Here, parameters have been chosen such that no Mott insulator at the center of
the trap is formed in the large $\gamma$ limit. The second scenario
corresponds to experiments in which the lattice depth is increased while the frequency of the
parabolic potential is kept fixed, Sec.IV B2.
In this case a unit filled Mott insulator is formed at the trap center and the inhibition of the
transport properties of the system is observed as the lattice
deepens.

\section{Single particle problem }
\subsection{Tight binding solution }

 The dynamics of a single  atom  in a one
dimensional  optical lattice plus a parabolic potential is described
by the Schr\"{o}dinger equation

\begin{eqnarray}
i\hbar \frac{\partial \Phi }{\partial t}=-\frac{\hbar ^{2}}{2m}\frac{
\partial ^{2}\Phi }{\partial x^{2}}+  V_{o}\sin^2
\left( \frac{\pi}{a} x \right)\Phi + \frac{m \omega_T^2}{2}x^2\Phi
,\label{Sch}
\end{eqnarray}

\noindent where   $\Phi(x)$ is the atomic wave function,
$\omega_T$ is the trapping frequency of the external quadratic
potential, $V_o$ is the optical lattice depth
 which is determined by the intensity of the laser beams,
$a=\lambda/2$ is the lattice spacing, $\lambda $ is the wavelength
of the lasers and  $m$ is the atomic mass.

If the atom is loaded into the lowest vibrational state of each
lattice well and the dynamics induced by external perturbations does
not generate interband transitions, the wave function $\Phi(x)$ can
be expanded  in terms of first-band Wannier functions only
\cite{mermin1976,Ziman1964}

\begin{equation}
\Phi (x,t)=\sum_{j}z_{j}(t)w_{0}(x-ja),\label{exp}
\end{equation}

\noindent where $w_{0}(x-ja)$ is the first-band Wannier function
centered at lattice site $j$, $t$ is time, and $\{z_j(t)\}$ are
complex amplitudes. For a deep enough lattice, tunneling to second
nearest neighbors can be ignored. This approximation known as the
tight binding approximation yields  the following equations of
motion for the amplitudes $\{z_j(t)\}$:

\begin{equation}
i\hbar \frac{\partial z_{j} }{\partial t}=-J\left( z_{j+1}+z_{j-1}\right) +%
\Omega j^2 z_{j},  \label{tba}
\end{equation}

\noindent with

\begin{eqnarray}
\Omega &=& \frac{1}{2}m a^2 \omega_T^2 ,\label{cha2om}\\
J&=&-\int dxw_{0}^{\ast }(x )H_{o}w_{0}(x-a)dx,\label{cha2J}\\
H_{o}&=&-\frac{\hbar ^{2}}{2m}\frac{%
\partial ^{2}}{\partial x^{2}}+  V_{o}\sin^2
\left(\frac{%
\pi }{a }x \right)
\end{eqnarray}

\noindent where $J$ is the tunneling matrix element between
nearest neighboring lattice sites. In Eq.(\ref{tba}) the overall
energy shift $\varepsilon _{o}$  given by
\begin{eqnarray}
\varepsilon _{o}&=&\int dxw_{0}^{\ast }(x)H_{o}w_{0}(x)dx,
\end{eqnarray}
\noindent has been  set to zero.

\subsection{Stationary solutions } \label{staso}

The stationary solutions of Eq.(\ref{tba}) are of the form
$z_{j}^{(n)}(t)= f_j^{(n)}e^{-i E_n t/\hbar}$, with $ E_n$ and
$f_j^{(n)}$ the $n^{th}$ eigenenergy and eigenstate, respectively.
Substitution into Eq.(\ref{tba}) yields

\begin{equation}
E_n f_j^{(n)}=-J\left(f_{j+1}^{(n)}+f_{j-1}^{(n)}\right) +%
\Omega j^2 f_{j}^{(n)} \label{stat}
\end{equation}

Equation (\ref {stat}) is formally equivalent to the recursion
relation satisfied by the Fourier coefficients of the periodic
Mathieu functions with period $\pi $. Therefore the eigenvalue
problem  can be exactly solved  by identifying the  amplitudes
$f_j^{(n)}$ and eigenenergies $E_n$  with the Fourier coefficients
and characteristic values of such functions , respectively
\cite{AS64}. In terms of Mathieu parameters the symmetric and
antisymmetric solutions are

\begin{eqnarray}
f_{j}^{(n=2r)} &=&\frac{1}{\pi }\int_{0}^{2\pi }ce_{2r}\left( x,-q\right) \cos (2jx)dx,  \quad  \\
f_{j}^{(n=2r+1)} &=&\frac{1}{\pi }\int_{0}^{2\pi }se_{2r}\left( x,-q\right) \sin (2jx)dx, \\
E_{n=2r} &=&\frac{\Omega }{4}a_{2r}\left( q\right) \quad \
E_{n=2r+1}=\frac{\Omega }{4}b_{2r}\left( q\right),
\label{nointhar}
\end{eqnarray}
\noindent with $r=0,1,2...$ and  $ce_{2r}(x,q )$ and $se_{2r}(x,q )$
the even and odd period $\pi $ solutions of the Mathieu equation
with parameter $q=\frac{4J}{\Omega}$ and characteristic parameter $%
a_{2r}(q )$ and $b_{2r}(q)$ respectively:
$\frac{d^{2}ce_{2r}}{dx^{2}}+(a_{2r}(q )-2q\cos (2x))ce_{2r}=0$
and $\frac{d^{2}se_{2r}}{dx^{2}}+(b_{2r}(q )-2q\cos (2x))se_{2r}=0
$.\\

Solutions of the single particle problem are entirely determined
by the parameter $q=\frac{4J}{\Omega }$, which is
proportional to the ratio of the nearest-neighbor hopping energy
 $J$ to the energy cost $\Omega$ for moving a particle from the central
site to its nearest-neighbor.
The eigenvalues and eigenstates of the
system  are in general complicated functions of
$q$. However, asymptotic expansions exist in
literature which can help unveil the underlying physics for
different values of $q$. Most of the asymptotic expansions have
been available for almost 50 years due to the work of Meixner and
Sch\"{a}fke \cite{Meixner}. In the remainder of this section we
introduce such asymptotic expansions and use them to describe the
physics of the system in the  tunneling dominated regime where $q
> 1$, which we call {\it high} $q$ regime, and the  $q
< 1$, or {\it low} $q$ regime.\\

\subsubsection{ High $q$ regime ($4J \gtrsim \Omega$)}
\label{highq}

Most  experiments have been developed in the parameter regime
where $q \gg 1 $. For example for ${}^{87}$Rb atoms trapped in a
lattice with $\lambda=810 $ nm, a  value of  $q \geq 10$ is
obtained for a lattice depth of  2 $E_{R}$  if $\omega_T< 2  \pi
\times  538 $ Hz and for a lattice of 50 E$_{R}$ if $\omega_T <2
\pi \times 7.5 $ Hz. Here $E_R$ is the photon recoil energy
$E_R=h^2/(2m\lambda^2)$, corresponding to a frequency $E_R/h=3.47$
kHz. Throughout this paper, in our examples we use atoms
with the mass of ${}^{87}$Rb.

In the high $q$ limit the periodic plus harmonic potential
possesses two different classes of eigenstates depending on their
energy. In fact, as we will show, eigenmodes can be classified as
low or high-energy depending on the quantum number $n$ being
smaller or larger than $2 \|\sqrt{q/2}\|$, respectively, where $\|
x \|$ denotes the closest integer to $x$. Physically, this
classification depends on which one of the two energy scales in
the system, the tunneling or the trapping energy, is dominant. To
the energy classification corresponds a classification based on
localization of the modes in the potentials. In particular, the
low-energy (LE) modes are extended around the trap center and
high-energy (HE) modes are localized on the sides of the
potential. The existence of localized and extended states has been
tested experimentally \cite{Pezze,Ott} and studied theoretically
\cite{Ruuska,polkovnikov0,Hooley,Rigol}. Here we show how the
asymptotic expansions of the Mathieu solutions  can be
used to characterize them quantitatively.\\

{\it {Low-energy modes} ($n\ll\sqrt{q}$) {\it in the high $q$ regime}}

In the LE regime the average hopping energy $J$ is larger than
$\Omega$. In this regime the eigenmodes have been shown to be
approximately harmonic oscillator eigenstates
\cite{Ruuska,polkovnikov0,Hooley,Rigol}. Asymptotic expansions
valid to describe LE eigenmodes in the high $q$ limit have been
studied in detail in Ref. \cite{math}, where the eigenstates are
described in terms of generalized Hermite polynomials. Here we
simply outline the basic results and refer the interested reader
to \cite{math} and references therein for further details.

The asymptotic expansions for the eigenmodes can
be written as
%\begin{widetext}

\begin{eqnarray}
f^{(n=2r)}_j \approx
A_n\exp{\left(-\xi^2\left(\frac{1}{2}+\frac{(3+2n)}{16\sqrt{q}}\right)+\frac{
\xi^4}{48\sqrt{q}}\right)}\notag\\
\cdot \sum_{k=0}^rh_k^{(r)}\xi^{2k}\left(1+\frac{(3k-k^2+10 k
r)}{24\sqrt{q}}\right)\label{eigve}\\
f^{(n=2r+1)}_j \approx
A_n\exp{\left(-\xi^2\left(\frac{1}{2}+\frac{(3+2n)}{16\sqrt{q}}\right)+\frac{\xi^4}{48
\sqrt{q} }\right)}\notag\\
\cdot \sum_{k=0}^r\tilde{h}_k^{(r)}\xi^{2k+1}\left(1+\frac{(7k-k^2+10 k
r)}{24\sqrt{q}}\right) \label{eigvo}
\end{eqnarray}
%\end{widetext}
\noindent with $\xi=j \sqrt[4]{\frac{4}{q}}$,
$h_k^{(r)}=\frac{(-1)^{r+k}2^{2k}2r!}{(2k)!(r-k)!}$,
$\tilde{h}_k^{(r)}=\frac{(-1)^{r+k}2^{2k+1}(2r+1)!}{(2k+1)!(r-k)!}$
and $A_n$ a normalization constant. Notice that the coefficients
$h_k^{(r)}$ and $\tilde{h}_k^{(r)}$ are related to the Hermite
polynomial $H_{n}(x)$ by the relations
$H_{2r}(x)=\sum_{k=0}^rh_k^{(r)}x^{2k}$ and
$H_{2r+1}(x)=\sum_{k=0}^r\tilde{h}_k^{(r)}x^{2k+1}$.

\noindent The eigenenergies are approximately given by

\begin{widetext}
\begin{eqnarray}
E^{low}_{n}\approx\frac{\Omega}{4}\left\{-2q+4
\sqrt{q}(n+\frac{1}{2})-
 \frac{(2n+1)^{2}+1}{8}-\frac{((2n + 1)^3
+ 3(2n + 1))}{ 2^7 \sqrt{q}} +
O\left(\frac{1}{q}\right)\right\}\label{enelo}
\end{eqnarray}

\end{widetext}

 \noindent If one neglects corrections of order $1/\sqrt{q}$ and higher in Eqs. (\ref{eigve}) and
(\ref{eigvo}) and keeps only the first two terms in Eq.
(\ref{enelo}), the expressions for the eigenmodes and eigenenergies
reduce to
\begin{eqnarray}
E_n=-\Omega q/2+\Omega\sqrt{q}(n+\frac{1}{2})\\
f^{(n)}_j\approx\sqrt{\frac{\sqrt{2}}{ 2^n n!\sqrt[4]{q \pi^2
}}}\exp\left(-\frac{\xi^2}{2}\right)
H_n\left(\xi\right).\label{eigv}
\end{eqnarray}

\noindent The above expressions correspond to  the  eigenvalues and
eigenenergies (shifted by $-\Omega q/2$)  of a harmonic oscillator
with an effective trapping frequency $\omega ^{\ast }$ and an
effective mass $m^{\ast }$. The effective frequency and mass are
given by
\begin{eqnarray}
\hbar \omega ^{\ast }&=&\Omega \sqrt{q}=\hbar \omega _{T}\sqrt{\frac{m}{m^{\ast }}},\label{effere}\\
 m^{\ast }
&=&\frac{\hbar ^{2}}{2Ja^{2}} \label{effema},
\end{eqnarray}
The harmonic oscillator character of the lowest energy modes in the
combined lattice plus harmonic potential is consistent with the fact
that near the bottom of the Bloch band the dispersion relation has
the usual free particle form with $m$ replaced by $m^*$. It is
important  to emphasize that the expressions for the
effective mass and frequency Eqs. (\ref{effere}) and  (\ref{effema})
are only valid in the tight-binding approximation.

 Higher order terms introduce corrections to these harmonic
oscillator expressions, that become more and more important as the
quantum number $n$ increases. These corrections  come from the
 discrete character of the tight-binding equation.
They can be calculated by replacing $ f_j^{(n)}$ with $f^{(n)}(\xi)$
and  taking the continuous limit of the hopping  term proportional
to $J$ in Eq.(\ref{stat}). This procedure yields:

 \begin{eqnarray}
\left( -\frac{1}{2} \frac{\partial^2 }{\partial
\xi^{2}}-\frac{1}{12{a_{ho}}^2} \frac{\partial^4 }{\partial
\xi^{4}}+\dots+ \frac{1}{2}
 \xi^2\right) f^{(n)} =\widetilde{E}_n f^{(n)} \label{difeq}
\end{eqnarray}

\noindent where  $\xi=j \sqrt[4]{\frac{4}{q}}\equiv \frac{j}{
a_{ho}}$ ,  $a_{ho}= \sqrt{\frac{\hbar }{m^{\ast }\omega ^{\ast }}}=
\left(q/4\right)^{1/4}$ is a characteristic length of the system in
lattice units, which can be understood as an  effective harmonic
oscillator length, and  $\widetilde{E}_n= (E_n+2J)/(\hbar
\omega^*)$. To zeroth order in $1/\sqrt{q}$, ($a_{ho}~\to~ \infty$),
the differential equation (\ref{difeq})  reduces to the harmonic
oscillator Schr\"{o}dinger equation. Higher order corrections given
in Eqs.(\ref{eigve}), (\ref{eigvo}) and (\ref{enelo}),   can be
calculated by treating the higher order derivatives in
Eq.(\ref{difeq}) as a perturbation.\\

{\it {High-energy  modes ($n\gg\sqrt{q}$) in the high $q$ regime}}

High energy modes are close to position eigenstates since  for
these states the  kinetic energy required for an atom to hop from
one site to the next  becomes insufficient to overcome the
potential energy cost \cite{Ruuska,polkovnikov0,Hooley,Rigol}. By
using asymptotic expansions of the characteristic Mathieu values
and functions, we obtain  the following expressions for the
spectrum and eigenmodes

\begin{figure}[tbh]
\begin{center}
\leavevmode {\includegraphics[width=3.0 in]{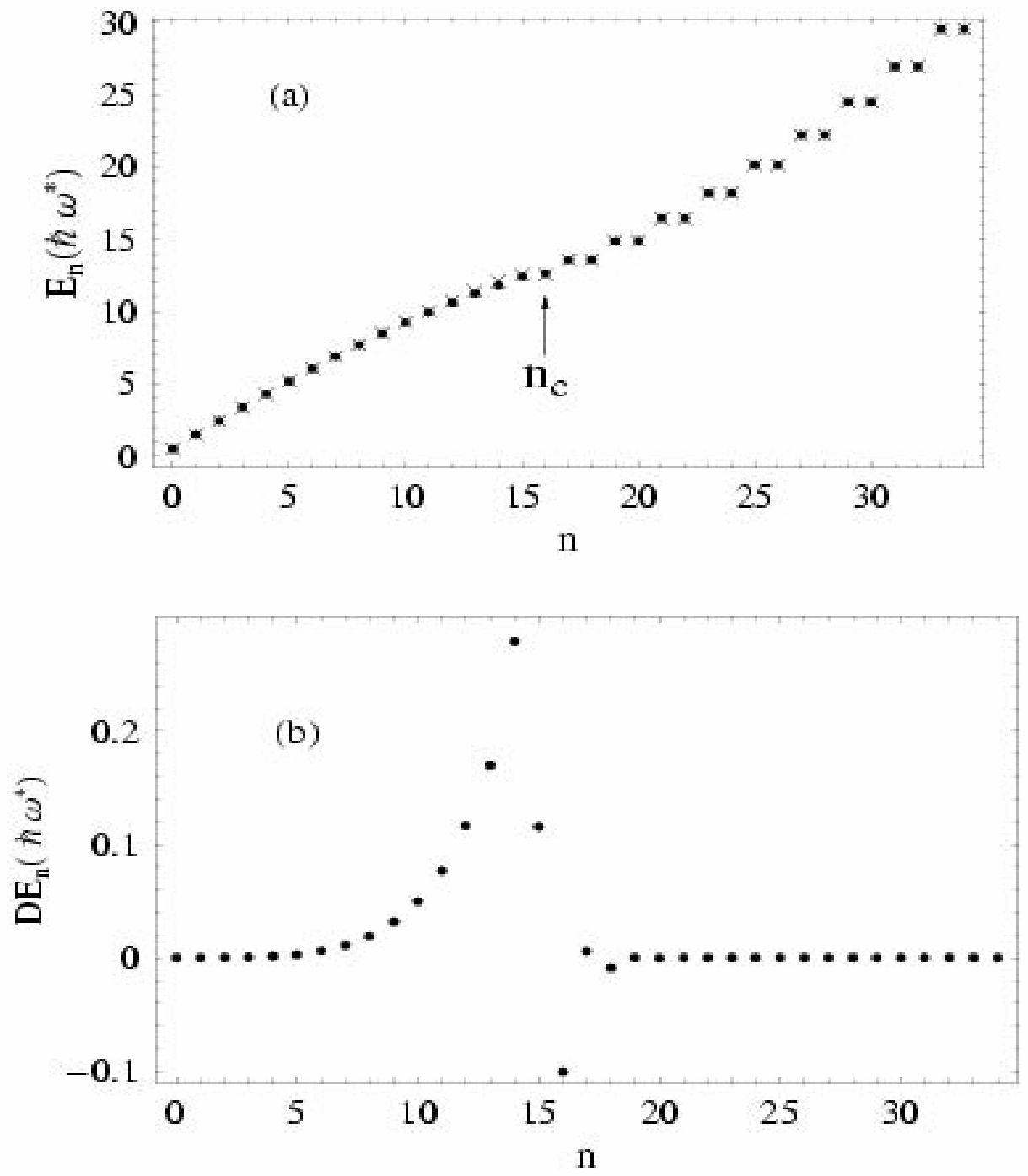}}
\end{center}
\caption{Upper panel (a): Spectrum of a particle in combined quadratic and periodic potentials as a function of the quantum number $n$. The
quadratic trap frequency and the
depth of the periodic potential are $\omega_{T}= 2\pi \times 60$ Hz and
$V_o=7.4 E_R$, respectively. Points are numerically obtained values,
while crosses are asymptotic expansions of the Mathieu characteristic
parameters. The arrow indicates the critical value $ n_c\approx 2 \|  \sqrt{q/2}\|$. Lower panel (b): Energy difference $DE_n$ between the numerically
obtained eigenvalues and the asymptotic expansions.
 }\label{harspe}
\end{figure}

\begin{figure}[tbh]
\begin{center}
\leavevmode {\includegraphics[width=3.0 in]{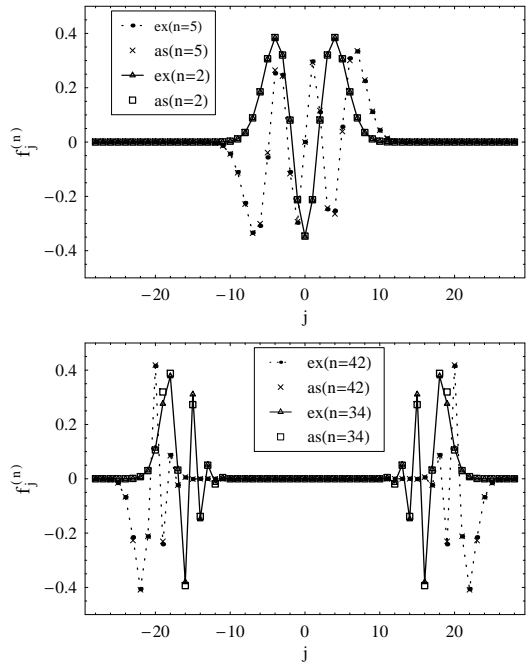}}
\end{center}
\caption{ Eigenmodes of a particle in a combined quadratic and periodic
potentials as a function of the lattice site $j$. The
 quadratic trap frequency and the
depth of the periodic potential are $\omega_{T}= 2\pi \times 60$ Hz and
$V_o=7.4 E_R$, respectively. Triangles and dots are numerically
obtained values ("ex" in the legend), while boxes and crosses are
asymptotic expansions of the Fourier coefficients of the periodic
Mathieu functions ("as" in the legend).
In the upper panel, triangles and boxes refer
to the $n=2$ mode, while dots and crosses refer to $n=5$.
In the lower panel, triangles and boxes refer
to the $n=34$ mode, while dots and crosses refer to $n=42$.
 }\label{harspe2}
\end{figure}

\begin{widetext}

\begin{eqnarray}
&&E^{high}_{n=2r} \approx E^{high}_{n=2r-1} \approx \frac{\Omega
}{4}\left( (2r)^{2}+ \frac{q^2}{2((2r)^{2}-1)}+ \frac{q^4\,\left(
7 + 5\,(2r)^2 \right) } {32\,{\left(  (2r)^2-1 \right)
}^3\left((2r)^2 -4   \right) } +\dots\right) \label{enehi}
\end{eqnarray}

\begin{eqnarray}
{f_j^{(high)}}^{n=2r}&&\approx {f_j^{(high)}}^{n=2r-1}\approx
A_n\left\{ \delta_{j,r}- \frac{q}{4}\left( \frac{\delta_{j, r-1}}{
2r-1} - \frac{\delta_{j,1 + r}}{1 + 2r} \right)  + \right.\notag\\
&& \left. \frac{q^2}{32}\left( \frac{\delta_{j, r-2}} {( 2r -2 )
( 2r -1 ) }-\frac{2( 1 + 4r^2 ) \delta_{j,r}} {{( 2r-1 ) }^2{( 1 +
2r ) }^2} + \frac{\delta_{j,2 + r}}{( 1 + 2r ) ( 2 + 2r ) }
\right)\right\}\pm \big\{j\rightarrow-j \big\} \label{modehi}
\end{eqnarray}

\end{widetext}
\noindent  with $A_n$ normalization constants. We note that
the asymptotic expansions Eqs.(\ref{enehi}) and (\ref{modehi})
for the HE-modes in the high $q$ regime are identical
to the asymptotic expansions for the modes in the low $q$ regime
reported below. While the latter are well known in literature,
to our knowledge they have never been used to describe
the high $q$ regime.
We actually found that as long as $n \gg \sqrt{q}$ these
asymptotic expansions reproduce reasonably well the exact results,
even for very large $q$. An example of this is given in Figs.
\ref{harspe} and \ref{harspe2}, which are discussed at the end of
this section.

The  high energy eigenstates are almost two-fold degenerate with
energy spacing mostly determined by $\Omega$. In Ref.
\cite{Hooley} the authors show that the localization of
these modes can be understood by means of a simple semiclassical
analysis. By utilizing a WKB approximation, the localization of
the modes can be  linked to the appearance of new turning points
in addition to the classical harmonic oscillator ones for energies
greater than $2J$. While classical harmonic oscillator turning
points are reached at zero quasimomentum, the new turning points
appear when the quasimomentum reaches the end of the Brillouin
zone and can therefore be associated with Bragg scattering induced
by the lattice.\\

{\it  {Intermediate states  ($n\sim\sqrt{q}$}) in the high $q$
regime }

In order to reproduce accurately the energy spectrum in the
intermediate regime one may expect that  many terms in the
asymptotic expansions have to be kept. To  estimate the energy
range where the spectrum changes character from low to high, we
solve for the smallest quantum number $n_c$ whose energy
calculated  by using the low energy asymptotic expansion is higher
than the one evaluated  by using  the  high energy expansion. That
is

\begin{equation}
E^{low}_{n_c-1}\geq E^{high}_{n_c},\label{match}
\end{equation}
\noindent where $n_c$ is required to be even.

 Solution of  Eq. (\ref{match}) gives

 \begin{equation}
 n_c\approx 2 \|  \sqrt{q/2}\|  , \label{sceq}
 \end{equation}
 \noindent where  $\| x \|$
 denotes the closest integer to $x$.

 By comparison with the numerically obtained eigenvalues, we actually
 find that  using  Eq. (\ref{enelo})  for $n<n_c-1$ and   Eq. (\ref{enehi})
for $n\geq n_{c}-1$  is enough to reproduce the entire spectrum
quite accurately. Moreover, since at $n_c$ the energy is
approximately given by $E_{n_c}\approx 2J$, our analytical
findings for the transition between harmonic oscillator-like and
localized eigenstates are in agreement with  the approximate
solutions found in \cite{polkovnikov0,Hooley} by using a WKB
analysis.

\bigskip

 In Figs. \ref{harspe} and \ref{harspe2} we compare the above   asymptotic
 approximations to exact numerical results for the eigenenergies
 and eigenfunctions.
The parameters  used for  the plots are those of a system  with  a lattice depth of $7.4
 E_R$ and quadratic trap frequency $\omega_{T}= 2\pi \times 60 $Hz.
These values correspond to $J= 0.0357 E_R$, $\Omega=0.0009 E_R$
and $q=157$. Figure \ref{harspe}, upper panel, shows the
 lowest 35 eigenenergies as a function of the quantum number $n$.
 The value $n_c$, which is equal to $16$ in this case, is indicated by an
 arrow. The crosses represent the asymptotic solutions, Eq.
 (\ref{enelo}) and (\ref{enehi}), and the dots the numerically
 obtained eigenvalues. On the scale of the graph there is essentially no
 appreciable difference  between the two solutions for the entire spectrum.
The difference between the the numerically obtained energies and
the asymptotic expansions is plotted in the lower panel of Fig.~\ref{harspe}.
In the upper panel of  Fig.\ref{harspe2},  asymptotic expressions
 for the LE eigenvectors $n=2$(boxes)
and $n=5$(crosses)  are compared to the numerically obtained
eigenmodes (triangles and dots respectively).  The modes clearly
exhibit an harmonic oscillator character, and the agreement
between the asymptotic and numerical solutions is very good. In
the lower panel, the $n=34$ and $n=42$ eigenstates belonging to
the region $n> n_c$ are depicted. These states are localized far
from the trap center. While the overall shape of the modes is well
reproduced by the asymptotic solutions, for the chosen values of
$n$ small differences between the  asymptotic (boxes and crosses
respectively) and  numerical solutions (triangles and dots
respectively) can be observed. As expected, the convergence of the
asymptotic expansion to the exact solution is better for  $n=42$
than for  $n=34$, as the former  has a larger value of $n$ than
the latter.

\subsubsection{ Low $q$ regime ($4J < \Omega$)}

This parameter regime is relevant for deep lattices. When $4J <
\Omega$ the kinetic energy required for an atom to hop from one
site to the next one is insufficient to overcome the trapping
energy  even at the trap center and all the modes are localized.
This is consistent with the previous analysis, because when $4J
\lesssim \Omega$, $n_c$ is less than one.  The asymptotic
expressions that
  describe this regime are \cite{AS64}

\begin{figure}
\begin{center}
\leavevmode {\includegraphics[width=3.0 in]{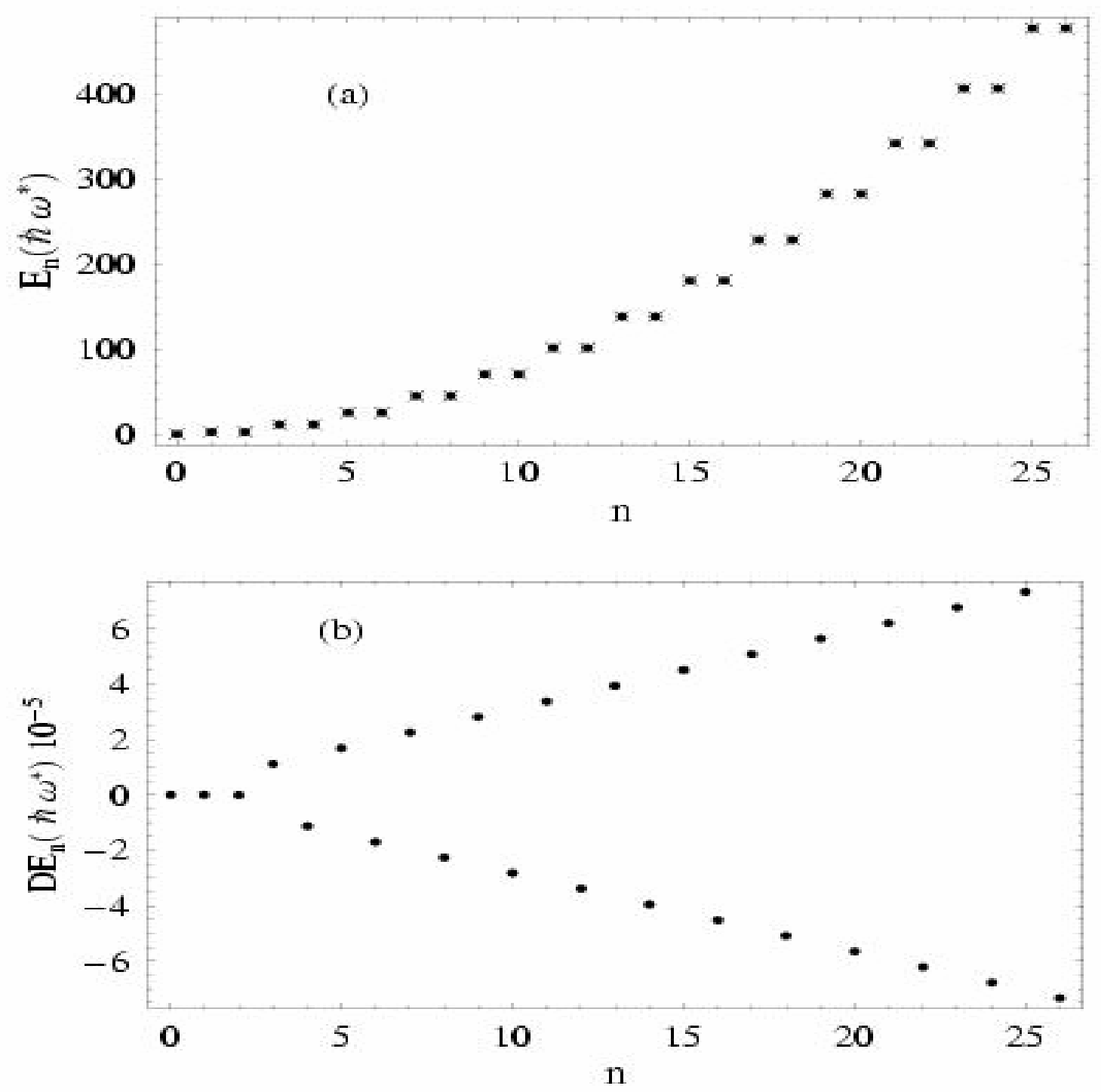}}
\end{center}
\caption{Upper panel (a): Spectrum of a particle in combined quadratic and periodic potentials as a function of the quantum number $n$. The
 quadratic trap frequency and the
depth of the periodic potential are $\omega_{T}= 2\pi \times 60$ Hz and
$V_o=50.0 E_R$, respectively. Points are numerically obtained values,
while crosses are asymptotic expansions of the Mathieu characteristic
parameters in the low $q$ limit. Lower panel (b): Energy difference $DE_n$ between the numerically obtained eigenvalues and the asymptotic expansions.
 }\label{harspe3}
\end{figure}

\begin{eqnarray}
&&E^{{\rm{Low}} q }_{n=0}\approx \frac{\Omega }{4}\left( -q^2
\frac{1}{2}+ q^4 \frac{ 7 } {128} +\dots\right)\notag\\
&&E^{{\rm{Low}} q }_{n=1}\approx \frac{\Omega }{4}\left(4 -q^2
\frac{1}{12}+ q^4 \frac{ 5 } {13824} +\dots\right)\notag\\
&&E^{{\rm{Low}} q  }_{n=2}\approx \frac{\Omega }{4}\left(4 + q^2
\frac{ 5}{12}-q^4 \frac{ 763 } {13824} +\dots\right)\notag\\
&&E^{{\rm{Low}} q }_{n=3}\approx \frac{\Omega }{4}\left(16 +q^2
\frac{1 }{30}- q^4 \frac{ 317 } {864000} +\dots\right)\notag\\
&&E^{{\rm{Low}} q  }_{n=4}\approx \frac{\Omega }{4}\left(16
+q^2\frac{ 1}{30}+ q^4 \frac{ 433 } {864000} +\dots\right)\notag\\
&&E^{{\rm{Low}} q }_{n\geqslant5}\approx  E^{high}_{n\geqslant5}
\end{eqnarray}

\noindent and

\begin{eqnarray}
{{f_j}^{{(\rm{Low}})}}^{n=0}&\approx&
A_0\left\{\frac{\delta_{j,0}}{\sqrt{8}}+ \frac{q}{\sqrt{8}}
\delta_{j, 1} +
\frac{q^2}{32}\frac{\delta_{j, 2} - \delta_{j,0}} {\sqrt{2}}\right\}\notag\\&&+ \big\{j\rightarrow-j \big\} \notag\\
{{f_j}^{{(\rm{Low})}}}^{n=1}&\approx&A_1\left\{ \delta_{j,1}+
\frac{q}{12}
 \delta_{j,2}+ q^2\left( \frac{\delta_{j, 3}}{384}-\frac{ \delta_{j,1}}
 {288}\right)\right\}\notag\\&&- \big\{j\rightarrow-j \big\} \notag \\
{{f_j}^{{(\rm{Low})}}}^{n=2}&\approx &A_2\left\{ \delta_{j,1}+
\frac{q}{12}
 \left(\delta_{j,2}-\frac{3}{2}\delta_{j,0}\right)+ \notag\right.\\&&
 \left.q^2\left( \frac{\delta_{j, 3}}{384}-19 \frac{ \delta_{j,1}}
 {288}\right)\right\}+\big\{j\rightarrow-j \big\} \notag \\
 {{f_j}^{{(\rm{Low})}}}^{n\geq3}&=&{f_j^{(high)}}^{n\geq3}
\end{eqnarray}
\noindent with $A_n$ normalization constants.
 In Fig. \ref{harspe3} the above expansions for the energies are  compared with the numerically
 calculated spectrum. Here the lattice is  $50 E_R$ deep and
the  external trap frequency is  $\omega_{T}= 2\pi \times 60 Hz$.
These lead to values of $J,\Omega$ and $q$ given by  $J= 2.9
\times 10^{-5} E_R$, $\Omega=0.0009 E_R$ and
 $q=0.13$. The asymptotic and numerical solutions perfectly agree on the scale of the graph. The energy difference between the numerically
obtained eigenvalues and the asymptotic expansions is plotted in Fig.~\ref{harspe3}, lower panel.

\section{ Center of mass evolution of a displaced system }
\label{dipole}

In recent experiments, the transport properties of one dimensional
Bose-Einstein condensates loaded in an optical lattice have been
studied after a sudden displacement of the quadratic trap
\cite{trey}. A strong dissipative dynamics was observed even for
very small displacements and shallow depths of the optical
lattice. This should be contrasted with previous experiments
performed with weakly interacting 3D gases where very small
damping of the center of mass motion was observed for small trap
displacements \cite{Cataliotti,Morsch}. Recent theoretical studies
have demonstrated that the strongly damped oscillations observed
in one dimensional systems reflect the importance of quantum
fluctuations as the dimensionality is reduced
\cite{polkovnikov1,polkovnikov2,polkovnikov3,julio}.

In this section we study the dipolar motion of ideal bosonic
and fermionic gases trapped in the combined lattice and harmonic
potentials. We start by writing an expression for the evolution
of the center of mass of an ideal gas with general quantum statistics, and then we use this expression
 to study the dipole oscillations for bosonic and fermionic systems. The simplicity of the
noninteracting treatment allows us to derive analytic equations
for the dipole dynamics for both statistics.

Later on, in section \ref{manybody}, we show how the
knowledge of  the bosonic and fermionic ideal gas dynamics
can be useful in describing  the dynamics
of the interacting bosonic system for a large range of
parameters of the trapping potentials.

\subsection{ Ideal gas dynamics}

Consider an ideal gas of $N$ atoms loaded in the ground state of
an optical lattice plus a quadratic potential initially displaced
from the trap center by $\delta$ lattice sites. The  initial state
of the gas  is described by a  mixed ensemble state with mean
occupation numbers determined by the appropriate quantum
statistics

\begin{eqnarray}
z_j(t=0)&=&\frac{1}{N}\sum_n  \overline{n}_n f^{(n)}_{j-\delta},\\
\overline{n}_n&=&\frac{1}{e^{\frac{E_n-\mu}{k_B T}}\pm1},\label{ave}
\end{eqnarray}

\noindent where $T$ is the initial temperature of the system, $k_B
$ is the Boltzmann constant, $\mu$ is the chemical potential which
fixes the total number of particles to $N$, $\sum
\overline{n}_n=N$, and the positive and negative signs in
Eq.(\ref{ave}) are for  fermions or bosons, respectively.

The time evolution of the center of mass of the gas is dictated by
the ensemble average

\begin{eqnarray}
\langle x(t)\rangle&=&\frac{1}{N}\sum_n \overline{n}_n \langle x_n
(t)\rangle\label{post}
\end{eqnarray}

\noindent with
\begin{eqnarray}
\langle x_n(t)\rangle&=& a \sum_{k,l} \left(c_l ^{(n)} c_k ^{(n)}
e^{-i
(E_k-E_l) t/\hbar} \sum_{j}j f_j^{(k)}f_j^{(l)*}\right) \nonumber\\
c_k ^{(n)}&=&\sum_j  f_{j-\delta}^{(n)} f_j^{(k)},
\end{eqnarray}

 \noindent where the quantities  $f_j^{(n)}$ and the energies
$E^{(n)}$ correspond to  eigenvalues and eigenenergies of the
undisplaced system. The coefficients $c_k ^{(n)}$  are  given by
the projection of the $n$ excited  displaced eigenstate onto
the $k$ excited undisplaced one.

Once the $E_n$ and $f^{(n)}_j$ are known, the center of mass evolution
can be calculated. In the following we discuss the zero
temperature dynamics  for  the ideal bosonic and fermionic  systems.

\subsubsection{Bosonic system}
\label{bosedam}

 At zero temperature  the bosons are Bose condensed and  $\overline{n}_n=
N \delta_{n0}$, where $\delta_{n0}$ is the Kronecker delta
function. The center of mass motion is then given by
\begin{eqnarray}
\langle x(t)\rangle&=& a\sum_{k,l} \left(c_l ^{(0)} c_k ^{(0)}
e^{-i
(E_k-E_l) t/\hbar} \sum_{j}j f_j^{(k)}f_j^{(l)*}\right) \nonumber\\
c_k ^{(0)}&=&\sum_j  f_{j-\delta}^{(0)} f_j^{(k)}.
 \end{eqnarray}

 If the initial displacement  of the atomic cloud is
small,  $ 2 \delta\ll n_c$, and the lattice is not very deep
($q\gg1$), localized eigenstates are initially not populated. Then,
only low-energy states are relevant for the dynamics  and the latter
can be modeled by utilizing the asymptotic expansions derived in
Sec. \ref{staso}. To simplify the calculations, we use the harmonic
oscillator approximation  for the eigenmodes (Eq.(\ref{eigv})), and
include up to  the quadratic corrections in $n$ in the
eigenenergies, which corresponds to keep the first three terms of
Eq.(\ref{enelo})). Even though this treatment is not exact, we found
that it properly accounts for the period and amplitudes of the
center of mass oscillations for small trap displacement. After some
algebra it is possible to show that the time evolution of the center
of mass is given by

\begin{equation}
\langle x\rangle=a \delta e^{-\left(\frac{\delta^2}{a_{ho}^2}
\sin^2\left(\frac{\Omega t}{8\hbar}\right)\right)}
\cos\left(\omega_o^*t- \frac{\delta^2 }{2a_{ho}^2}
\sin\left(\frac{\Omega t}{4 \hbar}\right)\right) \label{damsi}
\end{equation}

\noindent with $ \hbar \omega_o^*=  \hbar \omega^*-\Omega /4 $.

\begin{figure}[tbh]
\begin{center}
\leavevmode {\includegraphics[width=3.2 in]{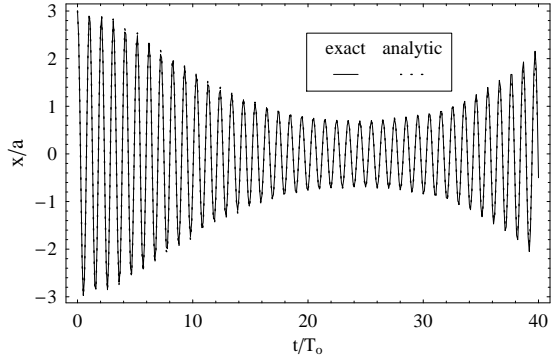}}
\end{center}
\caption{Center of mass motion in
lattice units as a function of time for an ideal bosonic gas. In
the plot,  $V_o=7.4 E_R$, $\omega_T=2\pi \times 60 Hz$ and $\delta=
3$. The time has been rescaled by $T_o= 2 \pi/\omega^*$.
The solid and dotted lines are the numerical
 and analytical solution Eq.~(\ref{damsi}), respectively.
 } \label{bose}
\end{figure}

In Fig.~\ref{bose} we plot the average center of mass position in
units of the lattice spacing $a$ as a function of time for an
ideal bosonic system of atoms with ${}^{87}$Rb mass.
The solid line is obtained by  numerically solving the
tight binding Schr\"{o}dinger equation, Eq.(\ref{tba}),
while the dotted line is the analytical solution Eq.(\ref{damsi}).
 For the plot we used $V_o=7.4 E_R$,
$\omega_T=2\pi \times 60 Hz$ and  $\delta= 3$.  The time is shown
in units of $T_o=2\pi/\omega^*$, a characteristic time scale. The
two solutions exhibit very good agreement for the times shown.

The modulation of the  dipole oscillations predicted  by
Eq.(\ref{damsi}) can be observed in the plot. At early times,
$t\ll\hbar/\Omega$, the amplitude decreases exponentially as $
\exp(-\Gamma_o t^2)$, with $\Gamma_o=\left(\frac{ \Omega \delta
}{8\hbar a_{ho}}\right)^2$, and the frequency  is shifted from
$\omega^*$ by $\frac{\Omega}{4\hbar}(1-\frac{\delta^2}{2
a_{ho}^2})$. The initial decay does not correspond to real damping
in a dissipative sense, as in a closed system the energy is
conserved.  The decay is just an initial modulation   and after
some time  revivals must be observed. Because in the large $q$
limit  $\omega_0^* \ll \Omega/\hbar$, the revival time is
approximately given by  $4 h/\Omega$.

It is a general result that the dipole oscillations of a
harmonically confined gas in absence of the lattice are undamped.
The undamped behavior holds  independently  of the temperature,
quantum statistics and interaction effects  (generalized Kohn
theorem \cite{Kohn}). Equation (\ref{damsi}) shows how this result does not
apply when the optical lattice is present even for an ideal Bose
gas.

Recent experimental developments have opened the possibility to
create a non-interacting gas for any given strength of the
trapping potentials \cite{feshbach1,feshbach2}.  The techniques
utilize Feshbach resonances for tuning the atomic scattering
length to zero. These developments should  allow  for the
experimental observation of the modulation of the dipole
oscillations of an ideal gas predicted in this section.

\begin{figure}[tbh]
\begin{center}
\leavevmode {\includegraphics[width=3.2 in]{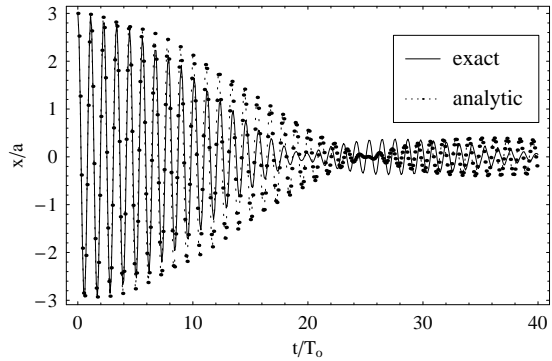}}
\end{center}
\caption{Center of mass motion in lattice units as a function of
time for $N=15$ fermions. Here $\omega_T= 2 \pi \times 20 Hz$ and
$V_o=7.4 E_R$ and $\delta=3$. The time is in units of
$T_o=2\pi/\omega^*$. The solid and dotted lines are the numerical
and analytical solution  Eq.~(\ref{postfin2}), respectively.
 }\label{fermit}
\end{figure}

\begin{figure*}
\begin{center}
\leavevmode {\includegraphics[width=7.0 in]{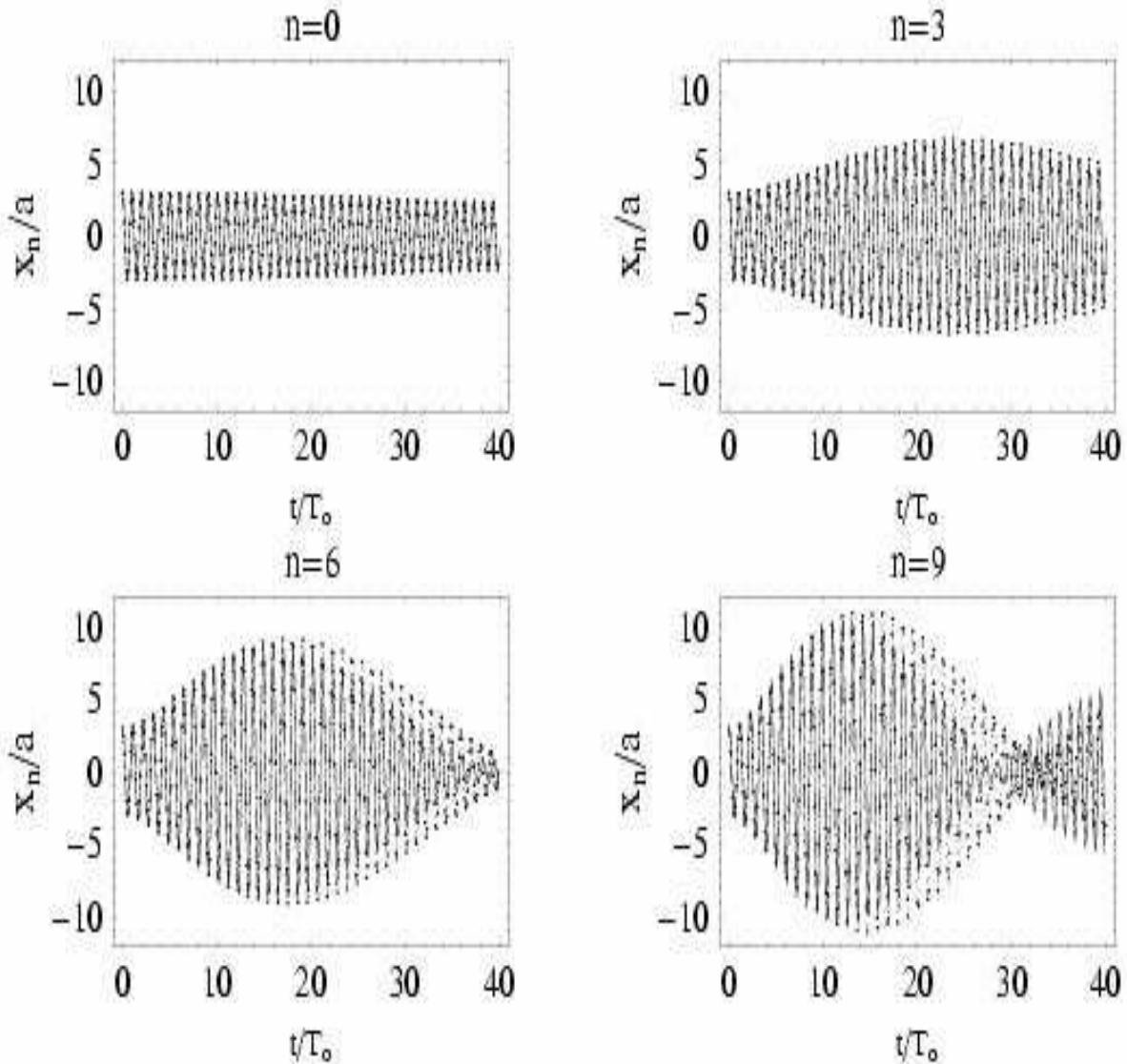}}
\end{center}
\caption{Center of mass motion in lattice units as a function
of time for some
initially occupied modes. The modes are labeled by the quantum number $n$.
As in Fig. \ref{fermit}, the parameters
are  $\omega_T= 2 \pi \times 20 Hz$, $V_o=7.4 E_R$, $N=15$,
$\delta=3$ and the time is in units of
$T_o=2\pi/\omega^*$. The solid line corresponds to the numerical
solution and the dotted line to the solution given by
Eq.~(\ref{postfin2}). When the center of mass evolution of  all
the different modes is added, from $n=0$ to $n=N-1$ one recovers
the total center of mass evolution shown in Fig. \ref{fermit}.
 }\label{fermic}
\end{figure*}

\subsubsection{ Fermionic system}
At zero temperature the  Pauli exclusion principle forces fermions
to occupy the lowest $N$ eigenmodes, and therefore $\bar{n}_n=1$ for
$0\leq n \leq N-1$ and zero elsewhere. The occupation of the first
$N$ displaced modes makes the condition of occupying only low-energy
eigenstates of the undisplaced potentials more restrictive than in
the bosonic case. Nevertheless, if the initial displacement, lattice
depth and atom number are chosen  such that only LE eigenstates are
initially populated, $\sum_{n=n_c-1}^\infty |c_n ^{(N-1)}|^2\ll 1$,
it is possible to derive simple analytic expressions for the dipole
dynamics. This is the focus of the remainder of this section.
Population of localized states considerably complicates the system's
dynamics and a numerical analysis is therefore required. This is
postponed to Sec.\ref{loc}.

When only LE undisplaced eigenstates are occupied,
 as explained for the bosonic system, to a good approximation
the eigenmodes can be assumed to be the harmonic oscillator
 eigenstates and only corrections quadratic
in the quantum number $n$ are relevant  in Eq.(\ref{enelo}). After
some algebra, the above approximations yield the following
expression for the time evolution of the center of mass:

\begin{widetext}

\begin{eqnarray}
\langle x(t)\rangle&=&  \frac{ 1 }{N} \sum_{n=0}^{N-1}
\langle x_n(t)\rangle \label{postfin}\\
\langle x_n(t)\rangle &\equiv& a \delta  {\rm{Re}}\left(
\exp\left\{i\omega_o^*t-\left(\frac{\delta^2(1-\chi(t))}{2
a_{ho}^2}\right)\right\}\tilde{x}_n(t)\right)\\
\tilde{x}_n(t) &\equiv&
\sum_{k=0}^{n-1}\left(\left(\frac{\delta}{a_{ho}} \right)^ {2k}
\frac{ (\chi(t)-1)^{2k} \chi(t)^{n-1-k}((n+1)\chi(t)+k-n)}{
(2k)!!(k+1)!}\prod_{s=1}^k(n+1-s)\right)+\frac{\left(\frac{\delta}{a_{ho}}\right)^{2n}(1-\chi(t))^{2n}}{(2n)!!}\label{postfin2}
\end{eqnarray}

\noindent where $\chi(t)=\exp^{(-i\Omega t/(4\hbar))}$ and
$\hbar\omega_o*=\hbar\omega^*-\Omega /4$.

\end{widetext}

The parameter $\chi(t)$ takes into account the quadratic
corrections  to the harmonic oscillator energies. The corrections
 are proportional to $\Omega /4$, and  due to the presence
of the lattice. In the limit $\chi(t)\rightarrow 1$,
$\tilde{x}_n(t)\rightarrow 1$ $\forall n$ and therefore the
amplitude of the dipole oscillations remains constant in time,
$\langle x\rangle=d\delta\cos(\omega^* t)$, as predicted by Kohn
theorem. The corrections quadratic in $n$ cause the modulation of
center of mass oscillations.

The modulation is caused not only by  the overall envelope
generated by  the exponential term
$\exp\left(-\delta^2(1-\chi(t))/(2 \tilde{a}_{ho}^2)\right)$,
which was also present in the bosonic case, but mainly from the
interference created by the different evolution of the $N$ average
positions $\langle x_n \rangle$ in the sum Eq.(\ref{postfin}). The
latter induces  a fast initial decay of the amplitude of the dipole
oscillations.

In Fig.\ref{fermit} we plot the center of mass motion of the
fermionic gas composed of $N=15$ atoms with the mass of $^{87}$Rb.
The solid and dotted lines
correspond to the numerical and analytic solutions, respectively.
Here the depth of the optical lattice is $7.4 E_R$, and $\omega_T=
2 \pi \times 20 Hz$. The amplitude of oscillation shows a rapid
decay in time. The analytic solution captures the overall
qualitative behavior of the numerical curve. Nevertheless,  only
at short times the agreement is quantitatively good. Population of
eigenstates which are not fully harmonic in character is
responsible for the disagreement at later times. This effect is
particularly relevant for the evolution of the displaced states
with larger quantum number, as explicitly shown in Fig.
\ref{fermic} where the time evolution of some displaced modes is
plotted. Again, the solid line is the numerical
solution and the dotted line is the analytic one. For the lowest
energy modes, $n=0$ and $n=3$, the agreement between the two
curves is almost perfect. For the higher energy modes $n=6$ and
$n=9$ the analytic solution is underdamped and overestimates the
collapse time.

Interestingly, the dynamics of the displaced excited modes
exhibits an initial growth of the amplitude. This behavior is a
pure quantum mechanical phenomenon due to the constructive
interference between the different phases of the undisplaced
eigenmodes during the  evolution. We explicitly checked for energy
conservation during the time evolution.  The amplitude increase is
captured  by the analytic solution and it allowed us to show that
the growth happens only when the ratio between the initial
displacement $\delta$ and $a_{ho}$ is less than one. While such a
behavior is not observable in the evolution of a fermionic cloud,
as the observable is the center of mass position  summed over all
initially populated modes $\langle x(t) \rangle$, the
experimental observation of growth for an  individual mode  may be
possible if an ideal bosonic gas is initially loaded in a
particular excited state, and then suddenly displaced.\\

As described above, the evolution of LE modes can be handled
analytically. On the other hand, when high-energy eigenmodes are
populated the dynamics is much more complicated. Nevertheless,
there is  another  simple  limiting case that can be actually
solved. This corresponds to the case when the displacement is large enough or the
lattice deep enough that the displaced  cloud has non-vanishing projection
amplitudes only onto high-energy undisplaced modes which can be
roughly approximated  by position eigenstates, $ f_j^{2r,2r-1}
\approx ( \delta_{j,r} \pm \delta_{-j,r} )/\sqrt{2}$. Then, one
finds  that $\langle x_n\rangle\approx a \delta $ for all $n$ and
thus $\langle x\rangle\approx a \delta$. That is, when only
high-energy eigenmodes are populated the dynamics is completely
overdamped and the cloud tends to remain frozen  at the initial
displaced position. We show later on in Sec. \ref{loc} where we
treat interacting atoms, that in the so-called {\it Mott}
insulator regime most populated modes are actually localized, and
this kind of overdamped behavior characterizes the dipole
dynamics.

\section{Many-body system }
\label{manybody}

\subsection{Spectrum of the BH-Hamiltonian in presence of an
external quadratic potential} \label{sspectrum}

The Bose-Hubbard ({\it BH}) Hamiltonian describes the system's
dynamics when the lattice is loaded such that only the lowest
vibrational level of each  lattice site is occupied \cite{Jaksch}
\begin{equation}
H_{BH}=\sum_{j} \left[ \Omega j^2 \hat{n}_j
-J(\hat{a}_j^{\dagger}\hat{a}_{j+1}
+\hat{a}_{j+1}^{\dagger}\hat{a}_{j})+\frac{U}{2}\hat{n}_j(\hat{n}_j-1) \right].\\
\label{EQNBHH}
\end{equation}
Here $\hat{a}_j$($\hat{a}^{\dagger}_j$) is the bosonic annihilation(creation) operator of a particle at site $j$ and
$\hat{n}_j=\hat{a}_j^{\dagger}\hat{a}_{j}$. $\Omega$ and $J$ are
defined as in Eqs. (\ref{cha2om}) and (\ref{cha2J}), while $U$ is
 an on-site interaction energy  given by $U=\frac{4 \pi
a_s\hbar^2}{m}\int d x|w(x_0)|^4$, with $a_s$ the s-wave
scattering length and $w_0(x)$  the first-band Wannier state
centered at the origin. The quantity $U$ is
the energy cost for having two atoms at the same lattice site.
The tunneling rate decreases for
sinusoidal lattices with the axial lattice depth $V_o$ as
\begin{equation}
 J= A  \left(\frac{V_o}{E_{R}}\right)^B  \exp \left(-C \sqrt{\frac{V_o}{E_{R}}}\right)
 E_R, \label{Jvo}
\end{equation}
\noindent  where the numerically obtained
constants are $A =1.397$, $B=1.051$ and $C=2.121$. The
interaction energy increases with $V_o$ as
\begin{equation}
U = \beta  E_R \left(\frac{V_o}{E_{R}}\right)^{1/4}, \label{Uvo}
\end{equation}
 \noindent where $\beta$ is a dimensionless constant proportional
to $a_s$. In current experiments, the one-dimensional lattice is
obtained by tightly confining in two directions atoms
loaded in a three-dimensional lattice. In this case $\beta=4
\sqrt{2 \pi}(a_s/\lambda) (V_{\perp}/ E_R)^{1/2}$, where
$V_{\perp}$ is the depth of the lattice in the transverse directions  \cite{anat,Olshanii}.
The parameter $\gamma=U/J$ therefore increases as a function of
the axial lattice depth as
\begin{equation}
\gamma(V_o) = \frac{\beta}{A}
\left(\frac{V_o}{E_{R}}\right)^{1/4-B} \exp \left(C
\sqrt{\frac{V_o}{E_{R}}}\right). \label{gvo}
\end{equation}

In the absence of the external quadratic potential, the bosonic spectrum
is fully characterized by the ratio between the interaction and
kinetic energies $\gamma$ and  the filling factor $N/M$,
where $M$ is the number of
lattice sites \cite{Fisher}. For $\gamma \ll 1$ and any $N/M$ ratio
the system is weakly interacting and superfluid. For $\gamma \gg 1$
and $M \geq N$ the  system fermionizes
 to minimize the inter-particle repulsion.
In this regime the bosonic energy spectrum mimics the fermionic one,
the correspondence being exact in the limit of infinitely strong
interactions. In particular, in a lattice model the onset of
fermionization is characterized by suppression of multiple particle
occupancy of single sites. This implies that fermionization
occurs for eigenstates whose energy is lower
than the interaction energy $U$. If
$M>N$ there are $M!/(N!(M-N)!)$ fermionized eigenmodes and the
dynamics at energies much lower than $U$ can be accounted for by
using these states only. On the other hand, if the lattice is
commensurately filled, $N=M$,  there is only one fermionized eigenstate,
and it corresponds to the ground state. All excited
states have at least one  multiply occupied site, and therefore
excitations are not fermionized. The
ground state corresponds to the Mott state with a
single particle per site and reduced number fluctuations. The
transition from the superfluid to the Mott state is a quantum phase
transition, and the critical point for one-dimensional unit filled
lattices is $ \gamma_c\simeq 4.65$ \cite{Fisher,Bat90}.

In the presence of the quadratic trap the spectrum is determined by
an interplay of $U$, $J$, $\Omega$ and $N$. In trapped systems the
notion of lattice commensurability becomes meaningless because the
size of the wave-function is explicitly determined by these
parameters. As a consequence, for any value of $N$ the ground state
can be made to be a Mott insulator with one atom per site at the
trap center by an appropriate choice of $U$, $J$ and $\Omega$
\cite{GAG}, and  the lowest energy modes can always be made to be
fermionized in the large $U$ limit. The purpose of this section is
to characterize fermionization and localization of the many-body
wave-function when both the quadratic and periodic potentials are
present, by relating the occurrence of the different regimes
to changes in the spectrum at low energies.\\

We performed exact diagonalizations of the {\it BH}-Hamiltonian
for $N=5$ particles and $M=19$ sites in presence of a quadratic
trap of frequency $\omega_T= 2\pi \times 150 $ Hz. For the
calculations, we chose $^{87}$Rb atoms with scattering length
$a_s=5.31$ nm, a lattice constant $a=405 $ nm, and therefore
$\Omega \simeq 0.0046 E_R$. We  fixed the transverse lattice
confinement to $V_{\perp} \simeq 25.5$ $E_r$ and varied the depth
of the optical lattice $V_o$ in the parallel direction from 2
$E_R$ to about 17 $E_R$. For these lattice depths, the energies
$U$ and $J$ both vary so that their ratio $\gamma$ increases from
$3.3$ to about $150$. Due to the changes in $J$, the ratio $q=4
J/\Omega$ characterizing the single-particle solutions decreases
from 130 to 4 with increasing $V_o$. The effective
harmonic-oscillator energy spacing $\hbar \omega^*$ decreases
approximately from 0.0525 $E_R$ to 0.0092 $E_R$. We used these
parameters because they are experimentally feasible and fulfill
the condition $U-\Omega ((N-1)/2)^2 > 0$ for the entire range of
the trapping potentials. Later on we discuss that, for deep enough
lattices, fulfillment of the last inequality ensures the existence
of an energy range in which eigenmodes are fermionized.

\begin{figure}
\begin{center}
\leavevmode {\includegraphics[width=3 in]{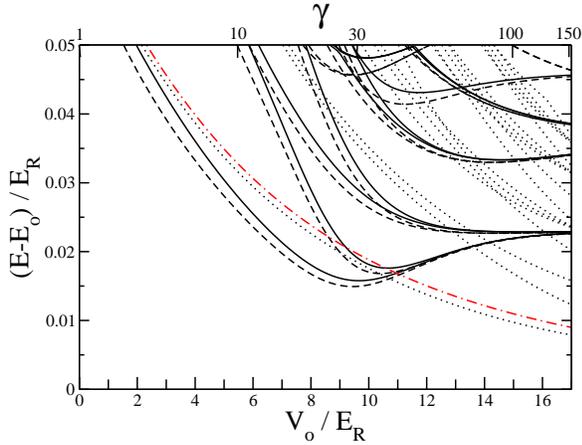}}
\end{center}
\caption{Energy spectra as a function of the depth of the axial
optical lattice $V_o$. The continuous, dotted and dashed lines
correspond to $N=5$ interacting bosons, non-interacting bosons and
fermions, respectively. The dashed-dotted line corresponds to
$\hbar \omega^*$. The horizontal axis on the top of the figure is
$\gamma=U/J$, and is only meaningful for interacting bosons. For
each energy spectrum, the corresponding ground-state energy $E_0$
has been subtracted.}\label{Spectrum}
\end{figure}

Figure \ref{Spectrum} shows the lowest eigenergies
of the {\it BH}-Hamiltonian as a function of $V_o$.
The continuous line is the exact solution for $N=5$ interacting bosons.
The dotted and dashed lines are the exact spectra for 5
non-interacting bosons and fermions, respectively. They have the same mass
and are trapped in the same potentials as the interacting bosons. Their spectra are shown for comparison purposes. For each spectrum the energy $E_0$ of the ground state has been subtracted.\\

In the absence of the optical lattice, $V_o=0$, the energy difference
$\Delta E_0=E_1-E_0$, or {\it energy spacing}, between the first excited and ground state equals
the harmonic oscillator level spacing $\hbar \omega_T$,
independent of statistics and interaction strength.
Figure \ref{Spectrum} show that this is no longer the case
in the presence of the lattice.

The dependence of $\Delta E_0$ on statistics is evident in the
plot, as the level spacing is different for ideal bosons and
fermions. In particular, the energy spacing for bosons is only
shifted from $\hbar \omega^*= \sqrt{4\Omega J}$ (dashed-dotted line)
by an amount which is almost constant for all lattice depths, while $\Delta E_0$ for
fermions clearly deviates from $\hbar \omega^*$, especially for
deep lattices. The behavior of $\Delta E_0$ in the two cases can
be understood by using the asymptotic solutions of the
single-particle problem. For ideal bosons $\Delta E_0$ is equal to
the energy difference between the first-excited and ground
single-particle eigenenergies. The ratio $q$ used for the plots is
such that the critical value $n_c$ of Eq.~(\ref{sceq}) is always
larger than 2, and therefore the ground and first-excited
eigenenergies are well described by Eq.~(\ref{enelo}). The
calculation of the energy difference using this equation yields
$\Delta E_0 \approx \hbar \omega^* - \Omega/4$. For fermions,
$\Delta E_0$ is equal to the difference between the energies of
the $n=N$ and $n=N-1$ single-particle excited states. For $V_o <
9.6 E_R$, the critical value $n_c$ is larger than $N$, and
therefore the energies of the $n=N$ and $n=N-1$ single-particle
excited states are also well described by Eq.~(\ref{enelo}). Then,
$\Delta E_0$ for fermions is smaller than for bosons because
lattice corrections are more important for higher quantum numbers,
and have all negative sign. On the other hand, for $V_o > 9.6
E_R$, $n_c$ is smaller than $N-1$ and  the energies of the $n=N$
and $n=N-1$ single-particle excited states are described by
Eq.~(\ref{enehi}). The transition of the single-particle
eigenmodes at the Fermi level from LE to HE  around $V_0 = 9.6
E_R$ is signaled by the minimum of $\Delta E_0$ for fermions. In
general, an estimate for the value of $J$ at which the minimum
takes place is

\begin{equation}
 J \sim \Omega ((N-1)/2)^2/2.\label{MottForm}
\end{equation}

\noindent This value is obtained by equating the Fermi energy
$E_{N-1}$, which is of order $\Omega ((N-1)/2)^2$ from
Eq.~(\ref{enehi}), to $E_{n_c}$, which is approximately $2 J$. The
transition of the single-particle eigenmode at the Fermi level
from LE to HE is also connected to the formation of a region of
particle localization at the trap center in the many-body density
profile. As explained in \cite{Rigol}, when $E_{N-1}$ is equal to
$E_{n_c}$ the on-site density in the central site of the trap
approaches 1 with reduced fluctuations. For $V_0 > 9.6 E_R$,
Fig.~\ref{Spectrum} shows that $\Delta E_0$ approaches an
asymptotic value $\Omega N$, value that can be derived from
Eq.~(\ref{enehi}). When the asymptotic value $\Omega N$ is
 reached, most
single-particle states below the Fermi level are
localized, and this yields a many-body density profile
with $N$ unit-filled
lattice sites at the trap center.

The dependence of the first excitation energy on interactions can
also be seen in Fig.~\ref{Spectrum}.
In fact, by comparing $\Delta E_0$ for the interacting bosons
to the value of $\Delta E_0$ for ideal bosons and fermions, three different
regimes can be considered: $1\lesssim \gamma \lesssim 10$, $10 \lesssim \gamma \lesssim
 30$ and $ \gamma > 30$. These regimes correspond
to the intermediate, fermionized-non-localized
and fermionized-Mott regimes, respectively. The weakly interacting regime
$\gamma \lesssim 1$ is only reached for $V_o \ll 2 E_R$ for our choice of
atoms and trapping potentials. For such lattice depths
the tight-binding approximation is not valid, and therefore
we do not show the spectra for this regime in Fig.~\ref{Spectrum}.
In the following we discuss the main features of the different regimes
focusing on the connection to the ideal bosonic and fermionic systems.

$\bullet$    For $\gamma \leq 1$ the interacting bosonic system is in the
\textit{weakly interacting} regime. In this regime the first
excitation energy is almost the same as the ideal bosonic one.
Most atoms are Bose-condensed,
interaction-induced correlations can be treated as a
small perturbation, and the spectrum can be shown to be well
reproduced by utilizing Bogoliubov theory \cite{anabog}.

$\bullet$ For $1< \gamma < 10$, the system is in the {\it
intermediate} regime, where $\Delta E_0$ for interacting bosons
deviates from the ideal bosonic energy spacing and approaches the
ideal fermionic one. Indeed, Fig.\ref{Spectrum} shows that for $
V_0 \lesssim 4 E_R$, $\Delta E_0$ for the interacting bosons lies
closer to the ideal bosonic energy spacing, while for $V_0 > 4$ it
lies closer to the ideal fermionic one. In the presence of the
optical lattice $\gamma$ increases exponentially with $V_o$, and
therefore the intermediate regime occurs for a relatively small
range of accessible trapping potentials, here for $ 1 \lesssim
V_o/E_R \lesssim 5.5$.

$\bullet$ For $\gamma \geq 10$
 the interacting spectrum approaches
the ideal Fermi spectrum and the system is in the
\textit{fermionized} regime. The numerical solutions show that the
energy difference between the energy spectra of interacting bosons
and fermions is of the order of $J^2/U$ and slightly increases for
larger frequencies of the quadratic trap.

In general, fermionization in the presence
of the external quadratic potential occurs for $N<M$
when the two following inequalities are satisfied
\begin{equation}
\gamma =U/J \gg 1 , U > \Omega ((N-1)/2)^2.\label{Req}
\end{equation}
While the first inequality is the same as for homogeneous lattices
and relates to the building of particle correlations, the second
inequality is specific to the trapped case and relates to
suppression of double particle occupancy of single sites. If the
interaction energy $U$ is larger than the largest trapping energy,
which corresponds to trapping an atom at position $(N-1)/2$, it is
energetically favorable to have at most one atom per well.
The average on-site occupation is therefore less or equal to one.

The second inequality in Eq.~(\ref{Req}) poses some limitations on the
choice of possible $\Omega$ and $U$ for a given number of trapped
particles. For our choice of the trapping potentials, this inequality
is satisfied for any $V_o$. Indeed, this is not an unrealistic assumption.
In recent experiments with ${}^{87}$Rb atoms, an array of fermionized gases
has been created with at most 18 atoms per tube \cite{Paredes}.
For such $N$, the condition $U > \Omega ((N-1)/2)^2$  can be fulfilled
for many different choices of experimentally feasible trapping
potentials.

\begin{figure}[tbh]
\begin{center}
\leavevmode {\includegraphics[width=3.0 in]{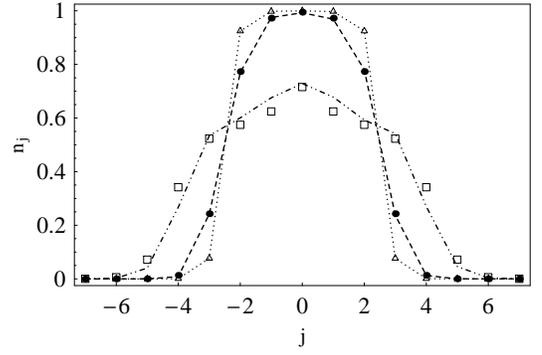}}
\end{center}
\caption{Density profiles  $n_j\equiv \langle \hat{n}_j\rangle$ as
a function of the lattice site $j$ for different lattice depths.
The dash-dotted,   dashed and  dotted lines correspond
to interacting bosons, while the  boxes, dots
and triangles correspond to ideal fermions for
$V_o/E_R=7, 12$ and $15$ lattice depths, respectively.
 }\label{feden}
\end{figure}

\begin{figure}[tbh]
\begin{center}
\leavevmode {\includegraphics[width=3.0 in]{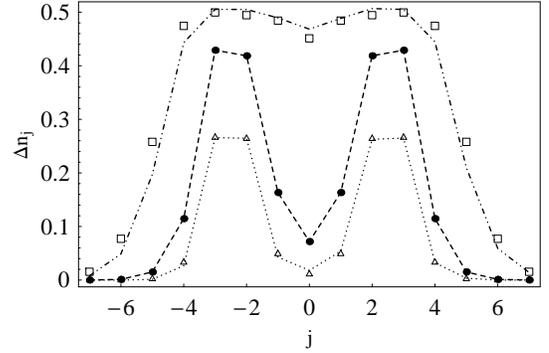}}
\end{center}
\caption{Number fluctuations $\Delta n_j \equiv
\sqrt{\langle\hat{n}_j^2\rangle-\langle\hat{n}_j\rangle^2}$ as a
function of the lattice site $j$ for different lattice depths.
Conventions and parameters are the same as in Fig.\ref{feden}
}\label{feflu}
\end{figure}

$\bullet$ For $\gamma > 30$ the system enters the fermionized-{\it
Mott} regime. In fact, Fig.~\ref{Spectrum} shows that for $\gamma
> 30$ the energy spacing for the interacting bosons  (and also for ideal fermions) begins to
increase until it reaches an asymptotic value at $\gamma \approx
150$.

The approaching of the asymptotic value signals the formation of a localized many-body state for the interacting bosons, the so-called Mott insulator state.
The relevant relation between $N$,
$\Omega$ and $J$ for the formation of an extended core of unit-filled sites
at the trap center with fluctuations mainly at the outermost occupied sites is

\begin{equation}
\Omega N \geq J.\label{Loc}
\end{equation}

This is explained as $\Omega N$ is the energy cost for moving
a particle from position $|(N-1)/2|$ to position $|(N+1)/2|$ at the
borders of the occupied lattice and it is the lowest
excitation energy deep in the Mott state.

Figure \ref{Spectrum} shows that for $\gamma \approx 150$ the
energies of the first four excited states become degenerate. This
degeneracy occurs because deep in the Mott regime the energy
required to shift all the atoms of one lattice site to the right
or left is the same as the energy required for moving an atom from
site $\pm (N-1)/2$ to site $\pm (N+1)/2$. For $\gamma \approx
150$, $J$ is approximately $\Omega$, and therefore the tunneling
energy is barely sufficient to overcome the potential energy cost
$\Omega$ for moving a particle from the central site of the trap to one
of its neighbors. In the single-particle picture,when $J\approx
\Omega$ all the single-particle states below $E_{N-1}$ are
 localized.\\

In order to better understand the formation of the Mott insulator, in
Figs.~\ref{feden} and \ref{feflu} the on-site particle number $n_j=
\langle \hat{n}_j \rangle $ and fluctuations $\Delta n_j =
\sqrt{\langle \hat{n}_j^2 \rangle - \langle \hat{n}_j \rangle^2}$
are plotted as a function of  the lattice site $j$, for different
lattice depths $V_o$. The lines and symbols are the results for
interacting bosons and ideal fermions, respectively. All the
$V_o$-values are such that the interacting bosonic system is
fermionized. This is mirrored by the overall good agreement between
lines and symbols for all the curves. The plots show that for $V_o=7
E_R (\gamma = 15$, dashed-dotted line), the largest average particle
occupation is $n_0\approx 0.7$, and number fluctuations are of the
order of $0.5$ in the central 9 sites, while for $V_o=12  E_R(\gamma
= 54$, dashed line),
 $n_0$ approaches 1 in the central 3 sites,
and fluctuations at the trap center drop to a value $\Delta n_j \approx
 0.1$. The sharp drop in particle fluctuations
clearly signals the localization at the trap center, and is in
agreement with $J<\Omega N$ for  $\gamma=54$. For $V_o=15
E_R(\gamma = 104$, dotted line), the Mott state is formed, as the
mean particle number in the five  central sites is one, with
nearly no fluctuations. Fluctuations are larger at lattice sites
far from the center, and due to the tunneling of particles to
unoccupied sites.
\\
Because of the small number of atoms that we use in the
calculations, $\Omega ((N-1)/2)^2 /2 $ and $\Omega N $ are of the
same order of magnitude. It is therefore not possible to clearly
distinguish the value of $J$ for which the on-site density at the
trap center approaches one from the value of $J$ for which the
Mott state is fully formed at the trap center, with reduced number
fluctuations in the central $N$ sites. Preliminary results
obtained with a Quantum Monte-Carlo code, Worm Algorithm
\cite{Proko}, confirm the existence of these two distinct
parameter regions when more atoms are considered, and therefore
the usefulness of both the energy scales
 $\Omega ((N-1)/2)^2 /2 $ and $\Omega N $ for interacting bosons.
These results will be published elsewhere.

\subsection{Many-body  dynamics}
\label{manydy}

In this section we study the  temporal dipole dynamics of an interacting
bosonic system composed of 5 atoms trapped in a combined quadratic
plus periodic confinement. The role of interactions on the
effective damping of the dipole oscillations is
studied by means of exact diagonalization of the Hamiltonian. As
discussed when dealing with  the ideal gas dynamics, such damping
is effective because it is due to dephasing and does not have a
dissipative character.

Assuming the system  is initially at $T=0$, the evolution of the
center of mass is given by:

\begin{eqnarray}
\langle x(t)\rangle&=& \sum_{l,k} A_{lk} e^{i \omega_{lk} t} \label{Prop} \\
A_{lk}&\equiv& a \sum_{j} j\langle \phi_l |
\hat{n}_j|\phi_k\rangle C_l^* C_k ,
\end{eqnarray}

\noindent with $\hbar \omega_{lk}\equiv E_l- E_k$ and where $E_l$
and $|\phi_l\rangle$ are eigenvalues and eigenmodes of the
{\it{BH}}-Hamiltonian.  The coefficients $C_l$ are the
projections of the initial displaced ground state
$|\varphi(0)\rangle$ onto the eigenfunctions $ \{ |\phi\rangle_l
\} $ of the undisplaced Hamiltonian, $C_l = \langle \phi_l
|\varphi(0)
 \rangle$.

For the exact evolution, the ground-state $|\varphi(0)\rangle$ is
calculated by  shifting the center of the quadratic trap by $\delta$
lattice sites.  The number of wells $M$ is 19 for all simulations,
which fixes the size of the Hilbert space to 33649. In the time
propagation we only keep those eigenstates whose coefficients $C_l$
are such that $|C_l|^2 > 10^{-3}$. The typical number of states
that fulfil this requirement is about 100. The accuracy of the
truncation of the Hilbert space during the time propagation has been
checked by increasing the number of retained states,
finding no appreciable changes in the dynamics.\\

We are interested in the dynamics both when a Mott insulator state
is not and is present at the trap center. These two cases are
discussed in Secs.\ref{nolo} and \ref{loc}, respectively. In
particular, in Sec. \ref{nolo} the interaction strength $U$ is
varied, while the ratio $q$, specifying the ideal gas dynamics, is
large and constant. For the chosen values of $q$ and $N$,
$J>\Omega N$ and for increasing $U$ the system fermionizes without
forming a Mott insulator at the trap center. In Sec. \ref{loc},
the dynamics of systems that do exhibit a Mott insulator in the
large $U$ limit is analyzed. In this case, $J$ and $U$ are
simultaneously varied by increasing the axial optical lattice
depth.

\subsubsection{Non-localized  dynamics}
\label{nolo} In the absence of the optical lattice, the equations
of motion for the center of mass are decoupled from those of the relative
coordinates. As only the latter are affected by
interactions, all modes excited during the collective oscillations
have the harmonic oscillator energy spacing $\hbar \omega_T$, and
therefore $\langle x(t)\rangle=\langle x(0)\rangle\cos(\omega_T t
)$.

When the lattice is present, the equations of motion for the
center of mass and relative coordinates are coupled, and thus the
many-body dynamics is interaction dependent. In
this section we fix $\omega_{T}=2 \pi \times 100 Hz$ and $V_o = 7
E_R$, and study the role of interactions in the many-atom dynamics
by varying $\gamma$ from $0$ to $200$, for constant $q=77$.
This can be experimentally realized
by tuning the scattering length of the system by means of Feshbach
resonances. In the following, we analyze the weakly interacting,
intermediate and
strongly interacting regimes separately.\\

{\it Weakly interacting regime: $\gamma \leq 1$}

\begin{figure}[tbh]
\begin{center}
\leavevmode {\includegraphics[width=3.0 in]{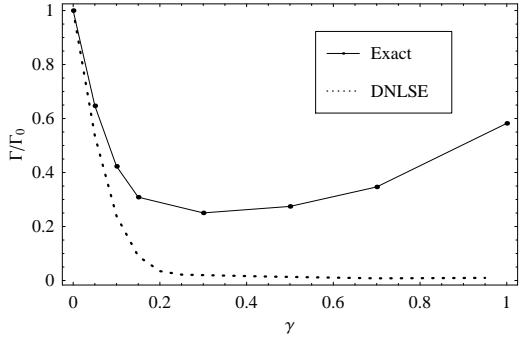}}
\end{center}
\caption{Effective damping constant of the dipole oscillations as a
function of $\gamma$ for a system in the weakly interacting regime.
Here, $q\approx 77$, $\omega_T = 100$ Hz and $\Gamma_o=
\left( \delta \Omega / 8 \hbar a_{ho}\right)^2$.
 }\label{wdamp}
\end{figure}

 In order to study the role of interactions  in the
weakly interacting regime, in Fig.~\ref{wdamp}  the effective
damping constant of dipole oscillations $\Gamma$ is shown as a
function of $\gamma$. The damping constant was calculated by fitting
the first 10 center of mass oscillations to the ansatz $\langle x(t) \rangle =a \delta \exp{(-\Gamma t^2)}\cos(\omega t)$, where $\Gamma$ and
$\omega$ are fitting parameters. This ansatz is chosen in analogy
to the non-interacting model. The effective damping $\Gamma$ is
in units of $\Gamma_o= \delta^2 \Omega^2 /(8 \hbar a_{ho})^2$,
which is the damping constant predicted by Eq.~\ref{damsi}.
The solid and dotted lines correspond to
$\Gamma$ as calculated by means of exact diagonalizations and by
numerically evolving the following {\it
Discrete Non-Linear Schr\"odinger Equation} (DNLSE) for the
amplitudes $\{z_{j}\}$
\begin{equation}
i\hbar\frac{\partial z_j}{\partial t} =-J(z_{j+1}+z_{j-1})+\Omega
j^2 z_j +U|z_{j}|^{2}z_{j}, \label{DNLSE}
\end{equation}

\noindent respectively. Eq.(\ref{DNLSE}) has been obtained by
replacing the field operator $ \hat{a}_{j}$  with the c-number
$z_{j}(t)$ in the Heisenberg equation of motion for $
\hat{a}_{j}$. Such replacement is justified for $\gamma \ll 1$ as
the many-body state is almost a product over identical
single-particle wave-functions. The amplitudes $\{z_{j}\}$ satisfy
the normalization condition $\sum_j|z_{j}|^{2}=N $. The initial
state used in the evolution of the DNLSE was found by numerically
solving for the ground state of Eq.~(\ref{DNLSE}), displaced by
$\delta=2$ lattice sites.

 In Fig.~\ref{wdamp}, the continuous and dotted
lines overlap for $\gamma \leq 0.05$, and show a decrease in the
damping constant with increasing interaction strength in the whole
range $\gamma \leq 0.2$. For values of $\gamma>0.05$ the mean
field and exact solutions start to disagree. While
 the exact solution has a minimum around  $\gamma \approx 0.2$,
and then grows for larger $\gamma$ values,
the mean field curve decreases monotonically to zero.

The fact that the mean-field solution decreases to zero for
increasing interactions is explained by noting that when
interaction effects become important the  density profile acquires
the form of an inverted parabola, or {\it{Thomas-Fermi}} profile,
$|z_j^{TF}(t) |^2\approx\Omega\left( ( j - \langle x(0)
\rangle/a)^2-R_{\rm{TF}}^2 \right)/U$, $\langle x(0) \rangle =a
\delta$. Substitution of the Thomas-Fermi profile in
Eq.~(\ref{DNLSE}) leads to the exact cancellation of the quadratic
potential, and thus, in the frame co-moving with the atomic cloud, the
atoms feel an effective linear potential. The spectrum of a linear
plus periodic potential is known to be equally spaced
\cite{Stark}, and therefore no damping due to dephasing is
expected.

It is important to note that the mean-field undamped oscillation
occurs only in a parameter regime far from dynamical
instabilities. In fact, as shown in previous theoretical and
experimental studies \cite{wu,Burger,smerzi}, when the initial
displacement is large enough to populate states above half of the
lattice band-width, mean-field dynamical instabilities induce a
chaotic dipole dynamics. In the framework of
this paper, this critical displacement corresponds to $\delta\approx n_c/2$.
For $\delta > n_c/2$, the initial ground-state
has a significant overlap with localized eigenmodes of the
undisplaced system, which are therefore populated during the
dipolar dynamics, causing damping. The importance of the population
of these modes is enhanced by the non-linear term, which
causes an abrupt suppression of the center of mass oscillations
at the critical point in the mean-field solution.\\

\begin{figure*}
\begin{center}
\leavevmode {\includegraphics[width=6.5 in]{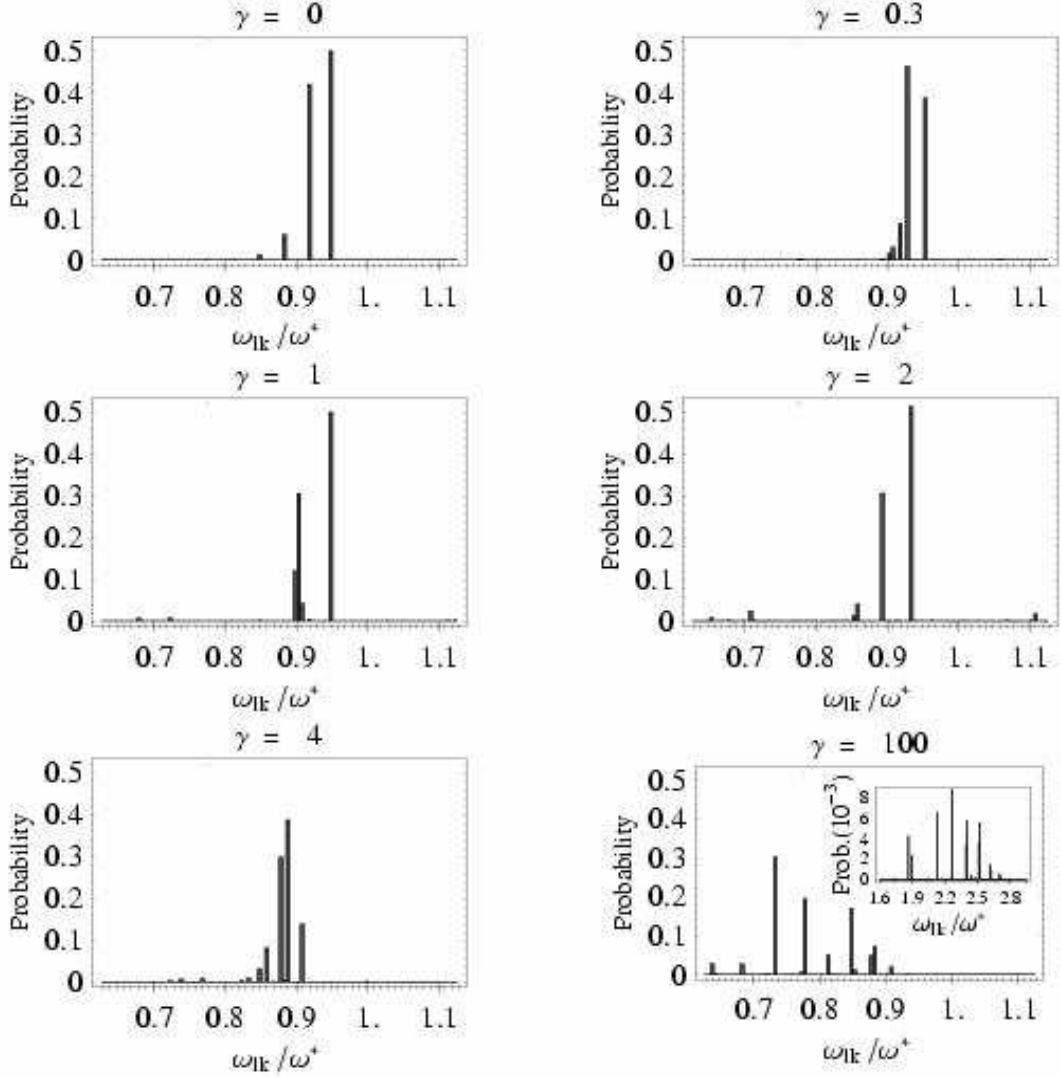}}
\end{center}
\caption{Probability distribution of frequencies  for different
values of $\gamma$, with $q\approx 77$ and $\omega_T=100$ Hz.
The frequencies $\omega_{lk}$ are in units of $\omega^*$.}\label{statis}
\end{figure*}
While the mean-field anaysis accounts for the decrease in the
damping constant, Fig.~\ref{wdamp} shows that it is not accurate for
$\gamma\gtrsim 0.2$. This is due to the fact that the mean-field
analysis completely neglects interaction-induced correlations.
These correlations are responsible for the quantum depletion of
the condensate, which causes  some atoms to be excited to
higher-energy single-particle eigenmodes which are
more affected by the lattice.
To show this effect in a quantitative manner,  in
Fig.\ref{statis} we plot the probability distribution of the
frequencies $\omega_{lk}$, Eq.~(\ref{Prop}), for some values of
$\gamma$. In the histograms, the height of a bar-chart centered
 at a given frequency $\omega$ is the occurrence probability  of $\omega_{lk}$.  The probability  is
  proportional to the normalized sum
over all the weight factors $|A_{lk}|$ whose frequency lies
between $\omega_{lk}-0.0025$ and $\omega_{lk}+0.0025$. The
frequencies are in units of the effective harmonic oscillator
frequency $\omega^*$ of Eq.~(\ref{effere}). This approach is
similar to the one used in Ref.~\cite{Lundh}, where the strength
function is used to study the collective dynamics induced by mono-
and dipolar excitations.

The histogram  for the case $\gamma=0$ shows a frequency
distribution with most frequencies in  the interval $0.85
\omega^*\leq \omega_{lk} \leq 0.96 \omega^*$. In particular, two
large peaks are observed in the range $0.9 \omega^*\leq
\omega_{lk}\leq 0.96\omega^*$. This is to be compared to the case
where the lattice is not present, and a single peak at $\omega^*$ is
expected. The observed frequency spread is due to the modification
of the harmonic oscillator spectrum introduced by the lattice and is
responsible for the observed damping in the ideal bosonic gas, as
explained in detail in Sec.~\ref{bosedam}.

 Figure \ref{statis} also shows that for $\gamma=0.3$ the
system has a  narrower frequency distribution. In this case
approximately $100 \%$ of the frequencies lie in the interval $0.9
\omega^* \leq \omega_{lk} \leq 0.96 \omega^*$. The frequency
narrowing from $\gamma=0$ to $\gamma=0.3$  is consistent with the
decrease in the damping constant shown in Fig.~\ref{wdamp}. For
larger values of $\gamma$, $\gamma=1$ and $2$, some modes with
frequency  smaller than $0.7 \omega^*$ and larger than $1.1
\omega^*$ start to contribute to the collective dynamics.
Population of such modes is related to the depletion  of the
condensate and signal the increased importance of quantum
fluctuations in the system.

Finally, we note that for our choice of $\Omega$, $N$ and $J$,
$\Omega((N-1)/2)^2 / J$ is approximately $0.2$, which is
the value of $\gamma=U/J$ at which the mean-field and exact solutions
begin to disagree. This suggests that the fulfillment of
the second inequality in Eq.~(\ref{Req}), $U>\Omega((N-1)/2)^2$,
is related to the failure of the mean-field approach,
even for $\gamma <1$.\\

{\it Intermediate Regime: $1<\gamma <10$}
\begin{figure}[tbh]
\begin{center}
\leavevmode {\includegraphics[width=3.5 in]{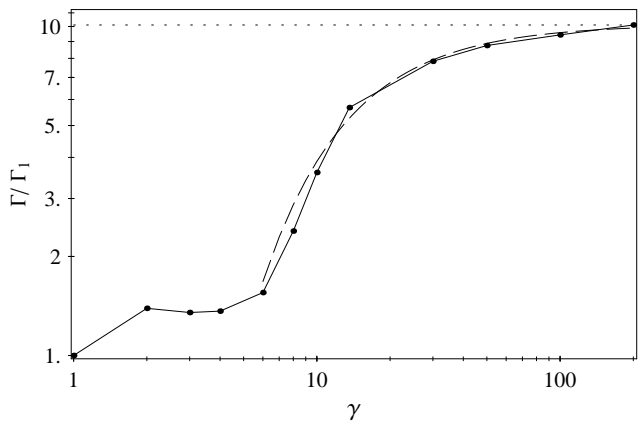}}
\end{center}
\caption{Effective damping rate of the dipole oscillations $\Gamma$
as a function of $\gamma$ in the intermediate and strongly correlated
regimes. Here, $q\approx 77$ and $\omega_T=100$ Hz.
The damping rate has been rescaled such
that $\Gamma(\gamma=1)=1$. The large and small dots
are for interacting bosons and non-interacting fermions, respectively.
The continuous line is a guide for the eye.
The dashed line is the best-fit curve
$\Gamma=10.13 \exp( -16.4 \gamma^ {-1.25})$ to the
exact damping rate for interacting bosons.}\label{idamp}
\end{figure}

In Fig.~\ref{idamp} the numerically obtained  damping constant is
shown for $\gamma>1$. In this parameter regime we find that
the function $a \delta \exp(-\Gamma t^2)\cos(\omega t)$ does not
provide a good fit to the center
of mass evolution. Instead, we find that a better fitting ansatz
is given by $a \delta \exp(-\Gamma t)\cos(\omega t )$ .
In the plot, the damping constant is normalized
to $\Gamma_1$, which is the damping rate at $\gamma=1$.

 We observe that for $2\leq \gamma < 5$ the damping is
almost constant.
This is consistent with the
fact that by inspection the spectrum of excited frequencies has a
similar shape and width in the entire transition region. An example
of such a frequency distribution is given in Fig.~\ref{statis}
for $\gamma=4$. The dominant peak is around
$\omega_{lk} \approx 0.9 \omega^*$, while multiple
peaks are noticeable between $0.7 \omega^* \leq
\omega_{lk} \leq 0.92 \omega^*$. The overall envelope of the
distribution has a long tail, as opposed to the case $\gamma=1$
where all the weight is roughly concentrated in just two
frequencies.

The increased  importance of the tails of the distribution for $\gamma > 1$
 qualitatively accounts for the transition from an exponential decay
quadratic in time  towards an exponential decay which  is linear
in time. In fact, for $\gamma < 1$ the probability distribution of
frequencies may be fitted by a Gaussian, while for $\gamma>1$
a better fit is provided by a Lorentzian-like profile.
 The Fourier transforms of such distributions give
precisely the observed functional form  of the decay of the dipole
oscillations. \\

{\it Strongly interacting regime $\gamma>10$}

In Fig. \ref{idamp} for $\gamma> 10$, the damping rate is shown
to rapidly increase and approach a finite
 asymptotic value which is depicted in the plot by
a dotted line. This asymptotic value $\Gamma_{\infty}$  corresponds to the damping
rate calculated for an ideal fermionic gas. The fermionic
damping rate is constant because here $J$ and
$\Omega$ are kept constant while $U$ increases.
The tendency to approach the fermionic damping rate as $\gamma$ increases is
a consequence of fermionization of the bosonic wave-functions
for $\gamma \gg 1$, Eq.~(\ref{Req}).
%explained by the fact that for $\gamma >10$ the system enters the
%fermionized regime. Actually, for our choice of parameters, the
%condition $U> \Omega (N-1)^2/4$ is satisfied already for
%$\gamma>0.2$ and $U/J \gg 1$ is the only  relevant constraint.
Numerically we find that the damping rate  approaches
$\Gamma_{\infty}$ exponentially,
$\Gamma(\gamma)=\Gamma_{\infty} \exp(-a \gamma^ \alpha)$, with a
best-fit exponent $\alpha$ of order $-1$. The fitting curve is
shown in the plot with a dashed line.

\begin{figure}
\begin{center}
\leavevmode {\includegraphics[width=3.5 in]{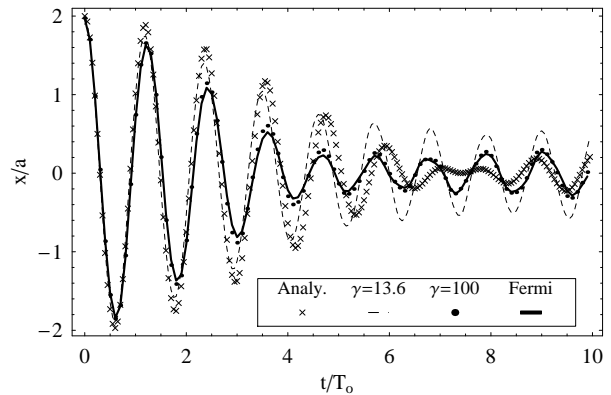}}
\end{center}
\caption{ Center of mass position in lattice units as a function
of time for interacting bosons, for $\gamma=13.6$(dashed line) and
$\gamma=100$(dots). The solid line is the center of mass position
for ideal fermions, while the crosses are the analytical
approximation to the fermionic solution Eq.(\ref{postfin2}). Here,
$q\approx 77$ and $\omega_T=2\pi \times 100$ Hz and $T_o=2
\pi/\omega^*$.
 }\label{fermitnm}
\end{figure}

The dipole dynamics of the bosonic and fermionic systems
are explicitly compared in Fig.~\ref{fermitnm},
 where we plot the  first 10 oscillations of the
center of mass after the sudden displacement of the trap.
 In the figure, the dashed line and the dots are the bosonic
 dynamics for $\gamma=13.6$ and 100, the solid line is the fermionic evolution,
and the crosses are the analytical
 approximation to the fermionic evolution Eq.~(\ref{postfin2}),
respectively. Consistent with Fig.~\ref{idamp}, we observe that
for increasing $\gamma$ the decrease of the amplitude of
oscillation for the bosons resembles more and more the one for
fermions. In particular, for $\gamma=100$, the curves for the
interacting bosons and ideal fermions nearly overlap. The
distribution of excited frequencies for $\gamma=100$ is shown in
Fig.~\ref{statis}. The frequency distribution is broad and
centered  around $\omega \approx 0.75 \omega^*$. Also  in the
inset small peaks are shown to  appear in the range $1.6 \omega^*
\leq \omega \leq 3 \omega^*$ (notice the different scale in the
inset). The broad distribution is due to the population of
single-particle states that are not  harmonic in character. For
the value of $q$ used for the calculations no single-particle
localized modes are occupied in the ground state before the trap
displacement. After the displacement about 90 percent of the atoms
occupy  non localized single-particle modes. The phase mixing
between these modes accounts for most of the observed damping. The
remaining 10 percent occupy localized states and the population of
these modes is responsible for the shift of the peak of the
distribution to lower frequency. In fact we show below,
Sec.~\ref{loc}, that a large population of localized states  with
$n \gg n_c$  yields a distribution which is peaked at
$\omega\approx 0$.

Finally, we note that the small population of localized modes
after the displacement  also explains why   the analytic solution
Eq.(\ref{postfin2}) reproduces the exact fermionic evolution in
Fig.~\ref{fermitnm} only qualitatively. In fact,
Eq.(\ref{postfin2}) was derived under the condition
$\sum_{n=n_c-1}|c_n^{(N-1)}|^2 \ll 1$, while here $n_c=12$  and
$\sum_{n=11}|c_n^{(N-1)}|^2\approx 0.1$.

\subsubsection{Localized dynamics}
\label{loc}
\begin{figure}
\begin{center}
\leavevmode {\includegraphics[width=3.5 in]{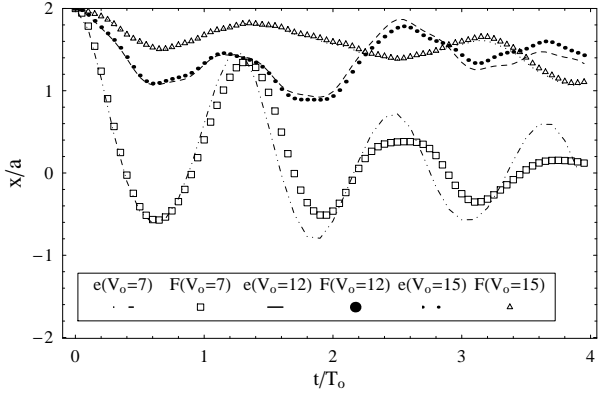}}
\end{center}
\caption{Center of mass position (in lattice sites) as a function
of time calculated for different lattice depths and a fixed
$\omega_T=2\pi \times 150$ Hz. The dash-dotted,  dashed and
small-dotted lines  are the exact solutions for interacting bosons
(e in the legend) and the boxes, large-dots and triangles are the
solutions for ideal fermions (F in the legend )  for $V_o/E_R=7,
12$ and $15 $, respectively. $T_o=2 \pi/\omega^*$.
 }\label{fermidam}
\end{figure}

\begin{figure}
\begin{center}
\leavevmode {\includegraphics[width=3.5 in]{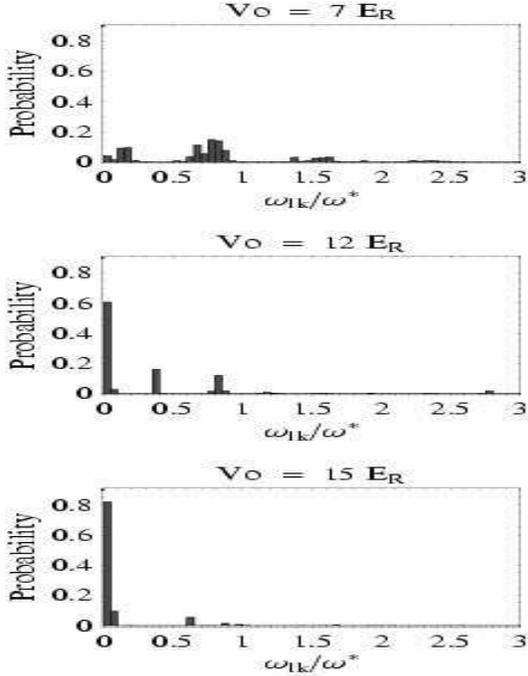}}
\end{center}
\caption{Probability distribution of frequencies for
$V_o/E_R=7,12$ and $ 15 $ (see text). The frequencies
$\omega_{lk}$ are given in  units of $\omega^*$. Because $\omega^*
\propto \sqrt{J}$, $\omega^*$ decreases with increasing lattice depth.
}\label{statcfe}
\end{figure}

In analogy to recent experiments
\cite{trey}, in this section we study the dipole dynamics when the depth $V_o$
of the optical lattice is varied,
 while the parabolic confinement is kept constant. Then, both $J$ and $U$ change as
 a function of the lattice depth, as explained in Sec.~\ref{manybody}.
  The parabolic confinement $\omega_T=2\pi \times 150 Hz$ has been chosen to be the same as
  in Sec.~\ref{manybody}, so that the energy spectrum exactly matches the one
discussed there, when $V_o$ is varied.

In Fig.~ \ref{fermidam} lines and symbols correspond to the time
evolution of the interacting bosons and ideal fermions,
respectively. In particular, the dashed-dotted, dashed and
small-dotted lines are for bosons,
 while boxes, large-dots and triangles are for
fermions with $V_o/E_R=7, 12$ and $15 $, respectively. For such
lattice depths $\gamma=14, 50$ and 100, respectively, and the
system is fermionized, as discussed in Sec.~\ref{sspectrum}. As
expected, the agreement between the bosonic and fermionic
solutions improves for larger $\gamma$-values, and is
  almost perfect for $\gamma=100$.

 Notably, for all the $\gamma$-values, no complete oscillation are observed,
 as the amplitudes of oscillations are strongly damped at very early times.
  The inhibition of the transport properties
in the experiment here envisioned is a direct  consequence of the
large population of single-particle states which are localized in
character. For the case  $V_o=7 E_R$ ($\Omega N/ J \approx 0.58$,
$\gamma=13.6$ , $q=134.2$ and $n_c\approx 8$)  before the
displacement the system is fermionized but non-localized.  On the
other hand, after the displacement about 20 percent of the atoms
occupy localized modes of the undisplaced potential. The
population of the localized modes with $n \gg n_c$ can be directly
linked to the presence of low-frequency peaks ($\omega_{lk}
\approx 0$) in the distribution of frequencies,
Fig.~\ref{statcfe}. Because 80 percent of the atoms occupy
non-localized  modes the center of mass position can still relax
to zero as shown in Fig.\ref{fermidam}. For the cases $V_o=12 E_R$
and $15 E_R$ the Fermi energy is larger than $E_{n_c}$, and even
before the displacement most states are localized. After the
displacement has taken place, about 60 and 90 percent of the atoms
occupy localized modes respectively, and the dynamics is
overdamped. This is mirrored by the appearance of a large peak at
$\omega_{lk} \approx 0$ in the probability distribution,
Fig.\ref{statcfe}, and
 by the fact that the center of mass position does not relax to
zero as shown in Fig.\ref{fermidam}.

\section{Conclusions}
We studied the spectrum and dipolar motion of  interacting and
non-interacting one-dimensional atomic gases  trapped in an
optical lattice plus a parabolic potential using the tight-binding
approximation.  We showed that the single-atom tight-binding
Schr\"{o}dinger equation  can be exactly solved by mapping it onto
the recurrence relation satisfied by  the Fourier coefficients  of
some periodic Mathieu functions. We used asymptotic expansions of
such functions  to fully characterize the eigenenergies and
eigenmodes of the system. Our analytic approach is complementary
to previous numerical and semiclassical analysis for single-atom
systems.  The advantage is that we could   explicitly calculate
the corrections to the harmonic oscillator spectrum introduced by
the lattice. The knowledge of these corrections allowed us also to
provide analytic expressions for the modulations of the  center of
mass motion  induced by the periodic potential when trapped ideal
bosonic and fermionic gases are suddenly displaced from the trap
center.

By means of numerical diagonalizations of the Bose-Hubbard
Hamiltonian we  studied the interacting many-body bosonic problem.
First, we characterized the changes in the low-energy excitation
spectrum  as a function of lattice depth, by comparing it with the
ideal Bose and Fermi spectra. Then, we stated the necessary
conditions for fermionization to occur and showed that it takes
place for a large range of experimentally accessible parameters.
We clarified the required conditions for the formation of a Mott
insulator at the trap center and linked its appearance to the
population of localized states at the Fermi level of the
correspondent ideal  fermionic system. We then studied the
many-body dipole dynamics and showed that, in the parameter regime
where the system is expected to be fermionized, the knowledge of
the single-particle solutions is  a powerful tool for the
understanding of  the strongly correlated  dynamics. By studying
the distribution of the frequencies pertaining  the many-body
modes excited during  the dipole  dynamics, we explicitly showed
the connection between the population of single-particle localized
states with  the inhibition of the transport properties of the
system. These spectral analysis allowed us also to gain some
insight into the dynamics in the weakly-interacting regime, where
an analysis beyond mean-field is required, and  in the complex
intermediate regime, where no mapping to single-particle solution
is possible.

\section{Acknowledgement}
We  thank Trey Porto and Eite Tiesinga for useful discussions. We
acknowledge financial support from an the Advanced Research and
Development Activity (ARDA) contract and the U.S. National Science
Foundation under grant PHY-0100767.

\bigskip
{\it{Note:}} We note that some of the points discussed here have
been just  recently pointed out in Ref.\cite{Rigol3} where the
authors discuss the dipole oscillations of 1D bosons in the
hard-core limit.

\end{document}